\documentclass[amsmat,amssymb,amsfonts,aps,prb,twocolumn,showpacs]{revtex4}
\usepackage{graphicx}
\usepackage{dcolumn}
\usepackage{bm}
\begin{document}

\title{Spin-phonon coupling in single Mn doped CdTe quantum dot}

\author{C. L. Cao (1,2), L. Besombes (2), J. Fern\'andez-Rossier(1)}
\affiliation{(1)Departamento de F\'{\i}sica Aplicada,
Universidad de Alicante, San Vicente del Raspeig, 03690 Spain \\
(2) Institut  Neel, CNRS, Grenoble, Avenue des Martyr, France}

\date{\today}

\begin{abstract}

The spin dynamics of a single Mn atom in a laser driven
CdTe quantum dot  is addressed theoretically. Recent
experimental results\cite{Le-Gall_PRL_2009,Goryca_PRL_2009,Le-Gall_PRB_2010}show that it is possible to induce Mn
spin polarization by means of circularly polarized optical
pumping. Pumping is made possible by the faster Mn spin
relaxation in the presence of the exciton.  Here we discuss
different Mn spin relaxation mechanisms.  First, Mn-phonon
coupling, which is enhanced in the presence of the exciton.
Second, phonon-induced  hole spin relaxation combined with
carrier-Mn spin flip coupling and photon emission  results
in Mn spin relaxation. We model the Mn spin dynamics under
the influence of a pumping laser that injects excitons into
the dot, taking into account exciton-Mn exchange and phonon
induced  spin relaxation of both Mn and holes.  Our
simulations account for the optically induced Mn spin
pumping.
\end{abstract}

\maketitle

\section{Introduction}
The tremendous progress in the miniaturization of
electronic devices has reached the point that makes it
crucial to address the effect of a single dopant in a
device and motivates the study of a single dopant spin to
store digital information\cite{Solotronics}. The
manipulation of a single atom spin in a solid state
environment has been demonstrated using several approaches,
like  scanning tunneling microscope on magnetic adatoms
\cite{Hirjibehedin_2007, Loth_2010}, or  optical probing of
NV centers in diamond \cite{Jelezko2004} and single
magnetic atoms in semiconductor quantum dots, the topic of
this paper. Single quantum dots doped with a single Mn atom
can be probed by means of single exciton spectroscopy in
photoluminescence (PL) experiments. This has been done both
in II-VI
\cite{Besombes_PRL_2004,Besombes2005,Leger05,Leger05B,Leger_PRL_2006,Leger_PRB_2007,Besombes_PRB_2008,Le-Gall_PRL_2009,Goryca_PRL_2009,Le-Gall_PRB_2010,Goryca_PRB_2010}
and III-V \cite{Kudelski_PRL_2007,Krebs_PRB_2009}
materials. In the case of single Mn doped CdTe dots,
information about the quantum spin state of a single Mn
atom is extracted from the single exciton quantum dot
photoluminescence due to the one on one relation between
photon energy and polarization and the electronic spin
state of the Mn atom
\cite{Fernandez-Rossier_PRB_2006,Qu05,Govorov04,Bhatta,Bhatta2007,Kuhn09,Kuhn11,JFR04,Joost08}.
This has made it possible to measure the spin relaxation
time of a single  Mn atom in a quantum dot under optical
excitation, using photon autocorrelation
measurements\cite{Besombes_PRB_2008}, and to  realize the
optical initialization and readout of the spin of the Mn
atom\cite{Le-Gall_PRL_2009,Goryca_PRL_2009,Le-Gall_PRB_2010}.

The observation of Mn spin orientation under quasi-resonant
optical pumping
\cite{Le-Gall_PRL_2009,Goryca_PRL_2009,Le-Gall_PRB_2010}
can be accounted for if {\em the Mn spin relaxation time is
shorter in the presence of a quantum dot exciton}
\cite{Govorov05,Le-Gall_PRB_2010,Cao10,Cywinski10}.  In
that situation, resonant excitation of  an optical
transition associated to a given Mn spin projection results
in the depletion of the laser driven Mn spin state, via Mn
spin relaxation in the presence of the exciton. Whereas
theoretical understanding of the exchange couplings between
electrons, holes and Mn spin in quantum dots permits to
account for the observed PL spectra
\cite{Besombes_PRL_2004,Leger05,Leger05B,Fernandez-Rossier_PRB_2006},
a complete understanding of the spin dynamics under the
combined action of laser pumping, incoherent spin
relaxation and coherent spin-flips  is still missing. In
this paper we make progress along this direction on two
counts. First, we discuss different Mn spin relaxation
mechanism, taking fully into account the interplay between
incoherent dynamics due to the coupling to a reservoir and
the coherent spin flips associated to exciton-Mn exchange
in the quantum dot. Our calculations show that the most
efficient Mn spin relaxation channel, in the presence of
the exciton, arises from a combination of phonon-induced
hole spin relaxation,  which turns the bright exciton into
a dark, followed by recombination to the ground enabled by
dark-bright mixing due to Mn-carrier spin flip exchange.
Thus, we provide a quantitative basis to a recently
proposed scenario\cite{Cywinski10}. Second, we model the Mn
spin dynamics with a rate equation for the Mn spin and the
Mn plus exciton spin states that includes the spin
relaxation rates between the few-body states calculated
from microscopic theory.

Our theory permits to model the experimental observations
and, importantly, it identifies the light-hole heavy-hole
mixing as a crucial parameter that determines not only the
PL lineshape
\cite{Leger05,Leger05B,Fernandez-Rossier_PRB_2006} but also
the amplitude of several spin-relaxation mechanisms at play
in this system.

The rest of this manuscript is organized as follows. In
section \ref{hamil0} we review the Hamiltonian for a single
Mn spin interacting with a single exciton in a quantum dot.
The anisotropic Mn-hole coupling is derived from a
simplified\cite{dot-holes,Fernandez-Rossier_PRB_2006}
single-particle description of the lowest energy quantum
dot hole states which affords  analytical expressions for
the critical parameter in the theory, the light-hole heavy
hole (LH-HH)  mixing \cite{dot-holes}. The  dependence of
the properties of the Mn-exciton states on the LH-HH mixing
are discussed. In section (\ref{Mn-phonon}) we discuss the
Mn spin relaxation due to Mn-phonon coupling, both with and
without an exciton in the quantum dot. Whereas this
mechanism is probably dominant for the Mn spin relaxation
in the optical ground state,  it is not sufficient to
account for the rapid Mn spin relaxation in the presence of
the exciton. This leads us to consider other spin
relaxation mechanisms.  In section (\ref{hole-spin}) we
describe the spin relaxation of holes due to their coupling
to acoustic phonons, using a Bir-Pikus Hamiltonian. Using
the simplified description of hole states, we  obtain
analytical results for the hole spin lifetime in a
non-magnetic dot,  which are in agreement with previous work using a  more
sophisticated description of single hole states  \cite{Woods04}.  We then compute
the lifetime of the exciton-Mn states due to hole spin
relaxation.  In section (\ref{dynamics}) we present our
simulations of the optical pumping process, using  rate
equations for the exciton-Mn quantum states, including the
laser pumping, the spontaneous photon emission, the Mn and
hole spin relaxation due to phonons.  Our simulations
account for the optical initialization and readout observed
experimentally.

\section{Exciton-$\rm Mn$ Hamiltonian}
\label{hamil0} In this section we describe a minimal
Hamiltonian model that can accounts for the PL spectra of
single Mn doped CdTe quantum dots. For that matter we need
to consider both the Mn spin in the unexcited crystal and
the Mn spin interacting with a quantum dot exciton. The
peaks in the PL spectra are associated to the energy
differences between the states of the dot with and without
the exciton.

\subsection{Mn spin Hamiltonian}
Mn is a substitutional impurity in the Cd site of  CdTe.
Thus, it has an oxidation state of $2+$,  so that the 5 $d$
electrons have spin $S=\frac{5}{2}$, resulting in a
sextuplet\cite{Furdyna} whose degeneracy is lifted by the
interplay of spin orbit and the crystal field. In an
unstrained  CdTe, the crystal field has cubic symmetry
which should result in a magnetic anisotropy Hamiltonian
without quadratic terms.  Electron paramagnetic resonance
(EPR) in CdTe strained epilayers\cite{Qazzaz} show that Mn
has an uniaxial term in the spin Hamiltonian. In a quantum
dot there could be some in-plane anisotropy as well, which
lead us to consider the following Hamiltonian:
\begin{equation}
{\cal H}_0= D M_z^2+ E(M_x^2-M_y^2) + g\mu_B \vec{B}\cdot\vec{M}
\label{H00}
\end{equation}
where $M_a$ are the $S=\frac{5}{2}$ spin operators of the
electronic spin of the Mn. The eigenstates of this
Hamiltonian are denoted by $\phi_m$
\begin{equation}
{\cal H}_0|\phi_m\rangle=E_m|\phi_m\rangle= E_m\sum_{M_z} \phi_m(M_z)|M_z\rangle
\end{equation}
where $|M_z\rangle$ are the eigenstates of $M_z$. In this
paper we neglect the hyperfine  coupling to the $I=\frac{5}{2}$
nuclear spin, which could affect the decay of the
electronic spin coherence \cite{Le-Gall_PRL_2009}. The
magnetic anisotropy parameters, $E$ and $D$ can not be
inferred from PL experiments, which are only sensitive to
the Mn-exciton coupling. EPR experiments\cite{Qazzaz} in
strained layer could be fit with $D=12\mu$eV,  $E=0$ and
$g=2.0$. Thus, the ground state should have $M_z=\pm \frac{1}{2}$,
split from the first excited state by $2D$. At 4 Kelvin and
zero magnetic field, we expect all the six spin levels to
be almost equally populated. We refer to these six states
as the ground state manifold , in contrast to the
excited state manifold, which we describe with 24
states corresponding to 4 possible quantum dot exciton
states and the 6 Mn spin states.

\subsection{Single particle states of the quantum dot}
We describe the confined states of the quantum dot with a
simple effective mass model. In the case of the conduction
band electrons, we neglect spin orbit coupling and we only
consider the lowest energy orbital, with wave function
$\psi_0(\vec{r})$, which can accommodate 1 electron with
two spin orientations.

In the case of holes, spin orbit coupling lifts the sixfold
degeneracy of the top of the valence band into a $J=\frac{3}{2}$
quartet and a $J=\frac{1}{2}$ doublet which, in CdTe, is more than
0.8eV below in energy.  Confinement and strain lift the
fourfold degeneracy of the $J=\frac{3}{2}$ hole states, giving rise
to a mostly $J_z=\pm \frac{3}{2}$ heavy hole doublet and a almost
$J_z=\pm \frac{1}{2}$ light hole doublet. Importantly, it is
crucial to include LH-HH mixing to describe the
experimental observation.

\subsubsection{Effect of confinement}

The top of the valence band states are  described in the kp
approximation with the so called Kohn-Luttinger Hamiltonian
\cite{Kohn-Luttinger,Broido-Sham}. For that matter, the
crystal Hamiltonian is represented in the basis of the 4
topmost $J=\frac{3}{2}$ valence states of the $\Gamma$ point. We
label them by $J=\frac{3}{2},J_z$. The resulting kp Hamiltonian can
be written as ${\cal H}_{\rm holes}={\cal H}_{\rm KL} $
 \begin{equation}
 {\cal H}_{\rm KL}= \sum_{i,j=x,y,z} V^{\rm KL}_{ij} J_i J_j k_i k_j  + \kappa\mu_BJ_z B
 \end{equation}
where $k_i$ are  the  wave vectors, $J_i$ are the spin
$\frac{3}{2}$ matrices, and $V^{\rm KL}_{ij}$ are coefficients
given in the appendix \ref{KL}.  The last term accounts for
the Zeeman coupling to an external field along the growth
direction.

The kp Hamiltonian for states in the presence of a quantum
dot confinement potential that breaks translational
invariance reads:

\begin{equation}
{\cal H}_{\rm kp}= -\hbar^2\sum_{i,j=x,y,z} V^{\rm KL}_{ij}
J_i J_j \partial_i \partial_j  + \kappa\mu_BJ_z B+
V(\vec{r}) \delta_{j_z,j_z'} \label{emass}
\end{equation}

In general, the numerical solution of equation
(\ref{emass}) can be very complicated.  Following previous
work\cite{dot-holes,Fernandez-Rossier_PRB_2006}, we make
two approximations that permit to obtain analytical
solutions. First, we model the quantum dot with a hard-wall
box-shape potential.  The dimensions of the box are $L_x$,
$L_y$ and $L_z$. This permits to write the wave function as
a linear combination of $|J=\frac{3}{2},J_z\rangle$ states
multiplied by the confined waves
 \begin{equation}
 \psi_{\vec{n}}(\vec{r})=\sqrt{\frac{8}{V}} Sin\left(\frac{n_x \pi x }{L_x}\right)Sin\left(\frac{n_y \pi y }{L_y}\right)Sin\left(\frac{n_z \pi z }{L_z}\right)
 \label{orbitalwave}
 \end{equation}
 Our second approximation is to restrict the basis set  to the fundamental mode  only, $n_x=n_y=n_z=1$. As a result, the quantum dot Hamiltonian reads
  \begin{equation}
 {\cal H}_{\rm kp}= -\hbar^2\sum_{i,j=x,y,z} V^{\rm KL}_{ij} J_i J_j \langle \partial_i \partial_j \rangle
 \label{dot}
 \end{equation}
where
  \begin{equation}
  \langle \partial_i \partial_j \rangle = \int \psi_{1,1,1}(\vec{r}) \partial_i\partial_j \psi_{1,1,1}(\vec{r})=
   \delta_{ij}  \left(\frac{2 \pi}{L_i}\right)^2
  \end{equation}
Thus, within this approximation, the quantum dot hole
states are described by a 4*4 Kohn Luttinger Hamiltonian
where the  terms linear in $k_i$ vanish  and the  $k_i^2$
terms are replaced by $\left(\frac{2 \pi}{L_i}\right)^2$.
The resulting matrix ${\cal H}_{\rm conf}$ has two
decoupled sectors denoted by $+$ ($J_z=+\frac{3}{2}$,$J_z=-\frac{1}{2}$)
and $-$ ($J_z=-\frac{3}{2}$,$J_z=+\frac{1}{2}$).  In the $(+\frac{3}{2}, -\frac{1}{2},
+\frac{1}{2}, -\frac{3}{2} )$ basis we have:
\begin{equation}
{\cal H}_{\rm conf}=\left(
\begin{array}{cc}
{\cal H}_{+} &  0 \\
 0&  {\cal H}_{-}
\end{array}
\right)
\label{conf1}
\end{equation}
with
\begin{eqnarray}
{\cal H}_+=\left(\begin{array}{cc}
\overline{P}+ \overline{Q} -\frac{3b}{2}&  \overline{R} \\
 \overline{R} & \overline{P}-\overline{Q}+\frac{b}{2}
 \end{array}\right)
 \label{KL-dotHH1}
\end{eqnarray}
and
\begin{eqnarray}
{\cal H}_-=\left(\begin{array}{cc}
 \overline{P}-\overline{Q}-\frac{b}{2}&  \overline{R} \\
 \overline{R} &\overline{P}+ \overline{Q} +\frac{3b}{2}
 \end{array}\right)
 \label{KL-dotHH2}
\end{eqnarray}
where $b\equiv \kappa\nu_B B$ and $\overline{P}$, $\overline{Q}$ and $\overline{R}$ are given in the appendix.
In order to find the corresponding energies and wavefunctions  it is convenient to write
these matrices as:
${\cal H}_{\pm} = a_{\pm} + \vec{h}_{\pm}\cdot\vec{\sigma}$
where $\vec{\sigma}$ are the Pauli matrices acting on the
pseudospin $\frac{1}{2}$ space of the $+$ and $-$ spaces,
$a_{\pm}=P\mp b/2$
and
\begin{equation}
\vec{h}_{\pm}=\left(\overline{R}, 0, \overline{Q}\mp b\right)= |\vec{h}_{\pm}| \left( Sin\theta_{\pm},0,Cos\theta_{\pm}\right)
\end{equation}
We keep only the two ground states (heavy-hole like),
denoted by  $|\Uparrow\rangle$ and $|\Downarrow\rangle$,
which are several meV away from the light-hole like states.
The ground state doublet for the  quantum dot  holes states
so obtained, neglecting strain,   can be written as
\begin{eqnarray}
|\Uparrow\rangle&=&  cos\frac{\theta_+}{2} |+\frac{3}{2}\rangle + sin\frac{\theta_+}{2} |-\frac{1}{2}\rangle \nonumber \\
|\Downarrow\rangle&=&  cos\frac{\theta_-}{2} |+\frac{-3}{2}\rangle + sin\frac{\theta_-}{2} |+\frac{1}{2}\rangle
\label{WF-KL-nostrain}
\end{eqnarray}
Thus, the LH-HH mixing parameters, $\theta_{\pm}$  depend
on the dot dimension, $L_i$, on the Kohn Luttinger
parameters, $\gamma_i$ and on the applied magnetic field
$B$.

\subsubsection{Effect of homogeneous strain}
We now consider the effect of the strain that arises from
the lattice  mismatch between the CdTe quantum dot and the
ZnTe substrate on the $J=\frac{3}{2}$ states of the valence band.
It has a similar effect that confinement, resulting in a
splitting of the $J=\frac{3}{2}$ manifold and a mixing of the  LH
and HH states. The Hamiltonian that describes the  effect
of strain, as described by the  strain tensor
$\epsilon_{ij}$, on the top of the valence band states in
zinc-blende semiconductors was proposed by Bir and Pikus.
We can  write the Bir and Pikus (BP) Hamiltonian as
\cite{Cardona-Yu} :
\begin{eqnarray}
{\cal H}_{\rm BP}= \left(a-\frac{9b}{4}\right) (e_{xx}+ e_{yy} + e_{zz})
+\nonumber\\
 +3 b \sum_{i=x,y,z}J_i^2e_{ii} + \sqrt{ 3}d \left(  \left(J_xJ_y+J_yJ_x
\right)e_{xy} + {\rm c.p.}\right)
\end{eqnarray}
where ${\rm c. p.}$ stands for cyclic permutation, and
$a=-3.4$eV, $b=-1.2$eV, $d=-5.4$eV  for
CdTe\cite{Cardona-Yu}.

For CdTe quantum dots grown in ZnTe,  we mainly consider
the effects of strain anisotropy in the growth
plane\cite{Leger_PRB_2007} and describe the strain by the
average values of $e_{xy}$ and $e_{xx}-e_{yy}$.  In this
approximation the BP Hamiltonian is reduced to a block
diagonal
  matrix in the $(+\frac{3}{2}, -\frac{1}{2}, +\frac{1}{2}, -\frac{3}{2} )$ basis:
 \begin{equation}
{\cal H}_{BP}=\left(
\begin{array}{cc}
{\cal H}_{BP+} &  0 \\
 0&  {\cal H}_{BP-}
\end{array}
\right)
\label{BP-4}
\end{equation}
where
 \begin{equation}
{\cal H}_{BP+}=\left(
\begin{array}{cc}
0 & \rho_s e^{-2i\varphi_s} \\
\rho_s e^{2i\varphi_s} & \Delta_{lh}
\end{array}
\right)
\end{equation}
\begin{equation}
{\cal H}_{BP-}=\left(
\begin{array}{cc}
\Delta_{lh} & \rho_s e^{-2i\varphi_s} \\
\rho_s e^{2i\varphi_s} & 0
\end{array}
\right)
\end{equation}
where $\Delta_{lh}$ is the HH-LH splitting, $\rho_s$ the
strained induced amplitude of the HH-LH mixing and $\phi_s$
the angle between the strained induced anisotropy axis in
the quantum dot plane and x (100) axis. Importantly, the
effect of confinement and the effect of strain have a very
similar mathematical structure. They both split the LH and
HH levels and mix them. The main difference lies in the
mixing term, which is real for the confinement Hamiltonian
controlled by the shape of the quantum dot and complex for
the BP Hamiltonian depending on the strain distribution in
the quantum dot plane.

\subsubsection{Combined effect of confinement and strain}
We finally consider the combined action of confinement and
strain described by ${\cal H}_{\rm holes}={\cal H}_{\rm
conf}+{\cal H}_{BP} $. Summing the Hamiltonians of
equations (\ref{conf1}) and (\ref{BP-4}) to obtain two
decoupled matrices for the $+$ and $-$ subspaces. They can
be written as
\begin{equation}
{\cal H}_{\rm tot,\pm} = A_{\pm} + \vec{H}_{\pm}\cdot\vec{\sigma}
\end{equation}
where $A_{\pm}=\overline{P}\mp\frac{b}{2}+\frac{\Delta_{lh}}{2}$
and
\begin{eqnarray}
\vec{H}_{\pm}=\nonumber\\
\left(\overline{R} + \rho_s cos(2\varphi_s), \pm \rho_s
sin(2\varphi_s) ,\overline{Q}\mp
b-\frac{\Delta_{lh}}{2}\right)
\end{eqnarray}
It is convenient to express the ground state doublet
associated to  ${\cal H}_{\rm holes}$ in terms of the
spherical coordinates  of the vectors $\vec{H}_{\pm}$,
$|\vec{H}_{\pm}|$, $\theta_{\pm}$ and $\phi_{\pm}$:
\begin{eqnarray}
|\Uparrow\rangle&=&Cos\frac{\theta_+}{2}|\frac{+3}{2}\rangle - Sin\frac{\theta_+}{2}e^{i\phi_+}|\frac{-1}{2}\rangle \nonumber\\
|\Downarrow\rangle&=&Cos\frac{\theta_-}{2}|\frac{-3}{2}\rangle - Sin\frac{\theta_-}{2}e^{i\phi_-}|\frac{+1}{2}\rangle
\label{WF-total}
\end{eqnarray}
where
\begin{equation}
e^{i\phi_\pm}=\frac{\overline{R}+\rho_s e^{\pm2i\varphi_s}}{|\overline{R}+\rho_s e^{\pm2i\varphi_s}|}
\end{equation}
Expectedly, this expression is formally very similar to that of equation (\ref{WF-KL-nostrain}).

Formally, we express eq. (\ref{WF-total}) as
\begin{equation}
|\sigma_h\rangle=\sum_{j_z} C_h(j_z) |j_z\rangle
\label{WF-general}
\end{equation}

Both in equations (\ref{WF-KL-nostrain}) and
(\ref{WF-total}) the ($+\frac{3}{2},-\frac{1}{2}$) sector is decoupled from
the ($-\frac{3}{2},+\frac{1}{2}$).  Whereas, this is not true in general,
it is sufficient to account for the correct symmetry of  a
variety of exchange couplings between the hole and both the
Mn and the electrons.

\subsection{Effective Mn-carrier exchange Hamiltonian}

\subsubsection{Hole-Mn Hamiltonian}

We now consider the exchange coupling of hole spin
($\vec{J}$) and the Mn spin ($\vec{M}$). The leading term
in the exchange interaction is the Heisenberg
operator\cite{Furdyna,Benoit}:
\begin{equation}
{\cal V}_{\rm exch}= \beta \delta(\vec{r}_h-\vec{r}_M)
\vec{J}\cdot\vec{M} \label{heis}
\end{equation}
where $\beta$ is the hole-Mn exchange coupling constant. For Mn in CdTe we have $\beta N_0=$0.88 eV, where $N_0$ is the volume of the CdTe unit cell\cite{Furdyna}. The
exchange interaction is taken as short ranged,  the Mn atom
is located at $\vec{r}_{Mn}$ and  $\vec{J}$ are the spin
$\frac{3}{2}$ angular momentum matrices. We represent the operator
(\ref{heis}) in the product basis  $|M\rangle\times
\sigma_h$. Thus, the exchange operator in the product basis
reads:
\begin{eqnarray}
\langle M|\langle\sigma_h|{\cal V}_{\rm
exch}|M'\rangle|\sigma_h'\rangle= \nonumber\\
\beta|\psi_0(\vec{r}_{Mn})|^2 \sum_a \langle
M|M_a|M'\rangle| \langle\sigma_h|J_a|\sigma_h'\rangle
\end{eqnarray}
\noindent where $\psi_0(\vec{r})$ is the envelope part of
the heavy hole wave function, eq. (\ref{orbitalwave}), and
$j_h\equiv \beta |\psi_0(\vec{r}_{Mn})|^2 $ is the hole-Mn
coupling constant, which depends both on a material
dependent constant $\beta$ and on a quantum dot dependent
property, the probability of finding the hole at the Mn
location.

After a straightforward calculation we obtain the effective
Mn-hole coupling spin model working in the space
$(M,\sigma_h)$ of dimension 12 as a function of the hole
wave function parameter $\theta$:
\begin{equation}
{\cal V}_{h-Mn} = j_{hx} M_x \sigma_x + j_{hy} M_y \sigma_y
+ j_{hz} M_z \sigma_z
\end{equation}

\noindent where the $j_{hi}$ are dimensionless coefficients
given, for $B=0$ and $\theta=\theta_+=\theta_-$ by:
\begin{eqnarray}
j_{hx}&=&\frac{j_h}{2}\left(\sqrt{3}Sin\theta+1-Cos\theta\right)\\
j_{hy}&=&\frac{j_h}{2}\left(\sqrt{3}Sin\theta-1+Cos\theta\right)\\
j_{hz}&=&\frac{j_h}{2}\left(1+2Cos\theta\right)
\end{eqnarray}

Notice that for $\theta=0$ there is no LH-HH mixing and we
have $j_{hx}=j_{hy}=0$ and $j_{hz}=\frac{3}{2}j_h$. In this
extreme case the Mn-hole coupling is Ising like and $M_z$
and $\sigma_z$ are conserved.  This limit is a good starting point
to model hole-Mn coupling in CdTe quantum dots\cite{Fernandez-Rossier_PRB_2006,Rossier-Aguado-PRL2007}

\subsubsection{Electron-Mn Hamiltonian}

In analogy to the hole-Mn bare coupling, the
electron-Mn coupling reads:
\begin{equation}
{\cal V}_{\rm e-Mn}= \alpha \delta(\vec{r}_e-\vec{r}_M) \vec{S}\cdot\vec{M}
\label{heis-elec}
\end{equation}
where $\vec{S}$ is the spin of the electron. Since the spin
orbit coupling has a very small effect on the $s$ like
conduction band, the effective exchange  for the quantum
dot electron and the Mn is also a Heisenberg  term given by
\begin{equation}
{\cal V}_{\rm e-Mn}= j_e \vec{S}\cdot\vec{M}=j_e\left(S_xM_z+S_yM_y+S_zM_z \right)
\label{heis-elec2}
\end{equation}
where $j_e\equiv\alpha |\psi_0(\vec{r}_{Mn})|^2 $ is the
electron-Mn coupling constant, which depends both on a
material dependent constant $\alpha$ and on a quantum dot
dependent property, the probability of finding the electron
at the Mn location. In our model we take the same orbital
wave function for the confined electron and hole, so that
the ratio $j_e/j_h=\alpha/\beta\simeq 4$ for CdTe. A
deviation from this expected ratio is indeed observed in
real CdTe quantum dots \cite{Besombes_PRL_2004}.

\subsection{Exciton-Mn wavefunctions and energy levels}

\subsubsection{ Hamiltonian} 
The effective Hamiltonian for the exciton in a single Mn
doped CdTe quantum dot is the sum of the single ion
magnetic anisotropy Hamiltonian, the Mn-electron and
Mn-hole exchange coupling and the electron-hole exchange
coupling
 \begin{eqnarray}
 {\cal H}= {\cal H}_S+ {\cal V}_{e-Mn}+{\cal V}_{h-Mn}+{\cal V}_{e-h}
 \label{HAMILTONIAN}
 \end{eqnarray}
 where
 \begin{eqnarray}
{\cal V}_{e-h}= j_{eh}S_z\sigma_h
 \end{eqnarray}
is the electron hole exchange coupling, neglecting
transverse components. Electron hole exchange is
ferromagnetic ($j_{eh}<0$) and   splits the 4 exciton
levels into two doublets, the low  energy dark doublet
($\Uparrow\uparrow,\Downarrow\downarrow $), denoted by
$X=\pm 2$ and the high energy  bright doublet
($\Uparrow\downarrow,\Downarrow\uparrow $) ($X=\pm 1$).

Since we consider two electron states
($S_z=\uparrow,\downarrow$), two hole states
($\sigma_h=\Uparrow,\Downarrow$), and six Mn states
$M_z=\pm \frac{5}{2}, \pm \frac{3}{2},\pm \frac{1}{2}$, the Hilbert space for the
Mn-exciton system has dimension 24.  Whereas we do obtain
the exact eigenstates of Hamiltonian (\ref{HAMILTONIAN}) by
numerical diagonalization,  it is convenient for the
discussion to relate them to eigenstate of the Ising, or
spin conserving part, of the Hamiltonian:
\begin{eqnarray}
{\cal H}= {\cal H}_{\rm Ising}+ {\cal H}_{\rm flip}
\label{HAMILTONIAN2}
\end{eqnarray}
where
\begin{eqnarray}
  {\cal H}_{\rm Ising}= DM_z^2 +  j_{eh}S_z \sigma_h +  j_{e} S_zM_z
 + j_{h} M_z \sigma_h
 \end{eqnarray}
 and
  \begin{eqnarray}
{\cal H}_{\rm flip}=E(M_x^2-M_y^2) + j_e \left(S_x M_x+S_y M_y\right) +\nonumber\\
+  \left(j_{hx}\sigma_x M_x+j_{hy}\sigma_y M_y\right)
 \label{flip}
 \end{eqnarray}
If we expand $j_{hx}$ and $j_{hy}$ in the series of LH-HH
mixing parameter $\theta$, they are the same in the first
order of $\theta$. For simplicity, we take
\begin{equation}
j_{h\perp}\equiv j_{hx}=j_{hy}=j_h\frac{\theta}{2\sqrt{3}}
\end{equation}
in the following calculation. In the case of a LH-HH mixing
induced by the anisotropy of the confinement described by a
hard-wall box shape potential, we get from the
Kohn-Luttinger Hamiltonian:

\begin{equation}
\theta=\frac{\pi^2\sqrt{3}\gamma_2|\frac{1}{L_x^2}-\frac{1}{L_y^2}|}{\sqrt{3\pi^4\gamma_2^2\left(\frac{1}{L_x^2}-\frac{1}{L_y^2}\right)^2+\gamma_1^2\left(\frac{-2}{L_z^2}+\frac{1}{L_x^2}+\frac{1}{L_y^2}\right)^2}}
\label{thetaLutt}
\end{equation}

\begin{figure}[t]
\begin{center}
\includegraphics[angle=0,width=0.9\linewidth]{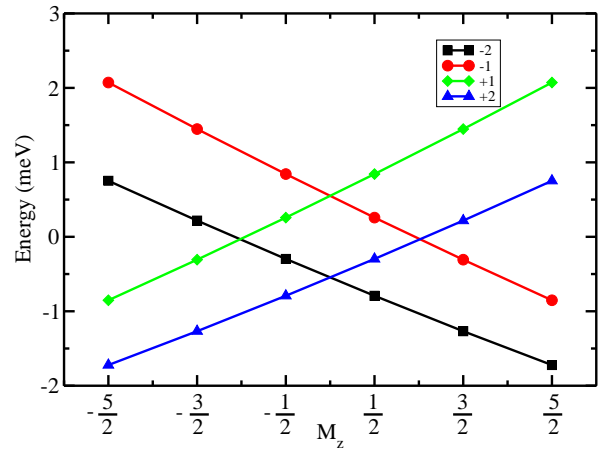}
\end{center}
\caption{ \label{fig1}(Color online) Scheme of the energy
levels of the quantum dot exciton interacting with 1 Mn
when spin-flip terms are neglected.  }
\end{figure}

The eigenstates of ${\cal H}_{\rm Ising}$ are trivially
given by the  product basis
  \begin{equation}
  |P\rangle\equiv
  |M_z\rangle|S_z\rangle|\sigma_h\rangle
  \label{product}
  \end{equation}
  with eigenenergies:
  \begin{equation}
 E_P
 = D M_z^2 +  j_{eh}S_z \sigma_h +  j_{e} S_z M_z
 + j_{h}M_z\sigma_h
  \end{equation}
Since the magnetic anisotropy term $D M_z^2$ is present
both in the ground state and exciton state manifolds, it
does not affect the PL spectra of the bright excitons.
Within this picture, for each of the 6 possible values of
$M_z$, there are 4 exciton states.   We  use a short-hand
notation to refer to the Ising states $P_{X}(M_z)$ where
$X=\pm 1,\pm 2$ labels the spin of the exciton, $X=S_z+
\sigma_z$. An energy diagram for the exciton levels, within
the Ising approximation, is shown in figure (\ref{fig1}).

The PL spectra of a single Mn doped quantum dot predicted
by the model of Ising excitons, ie, neglecting the spin
flip transitions, features 6 peaks corresponding to
transitions conserving $M_z$.  For the recombination of
$\sigma^+$ excitons ($S_z=-\frac{1}{2},\sigma_h=\Uparrow$) the high
energy peak corresponds to $M_z=+\frac{5}{2}$ and the low energy
peak to $M_z=-\frac{5}{2}$ on account of the antiferromagnetic
coupling between the hole and the Mn.  In the case of
$\sigma^-$ excitons the roles are reversed, but the PL
spectrum is identical at zero magnetic field.

\subsubsection{Wave functions}

When spin-flip terms are restored in the Hamiltonian, the
$P$ states are no longer eigenstates, but they form a very
convenient basis  to expand the actual eigenstates of
${\cal H}$, denoted by $ |\Psi_n\rangle$:
     \begin{equation}
 | \Psi_n\rangle=
  \sum_{P} \Psi_n(P) |P\rangle = \sum_{X,M_z}  \Psi_n(X,M_z) |X,M_z\rangle
  \label{exciton-vs-P}
  \end{equation}
In most cases, there is a strong overlap between $\Psi_n$
and a single state  $|P\rangle$. This is expected for
several reasons. First,  the single ion in plane anisotropy
is probably much smaller than the uniaxial anisotropy,
$D\gg E$. Second, the   electron-hole exchange, which is the
exchange energy in the system,  splits the dark and bright
levels. Thus, both  electron and hole spin flip due to the
exchange with the  Mn spin is  inhibited because they involve
coupling between energy split  bright and dark  excitons.
In addition, the electron Mn exchange is smaller than the
hole Mn exchange, whose spin-flip part is proportional to
the LH-HH mixing and approximately 10 times smaller than
the Ising part. In order to quantify the degree of spin
mixing of an exact exciton state $\Psi_n$, we define the
inverse participation ratio:
 \begin{equation}
 IPZ_n\equiv \sum_P |\Psi_n(P)|^4
 \end{equation}
This quantity gives a measure of the delocalization of the
state $\Psi_n$  on the space of  product  states of eq. (\ref{product}).  In the absence of mixing of different $P$ states, we have $IPZ_n=1$. In the case of a state equally
delocalized in the 24 states of the $P$ space, we would
have $\Psi_n(P)=\frac{1}{\sqrt{24}}$ and $IPZ_n=
\frac{1}{24}$.

\begin{figure}[t]
\begin{center}
\includegraphics[angle=0,width=1\linewidth]{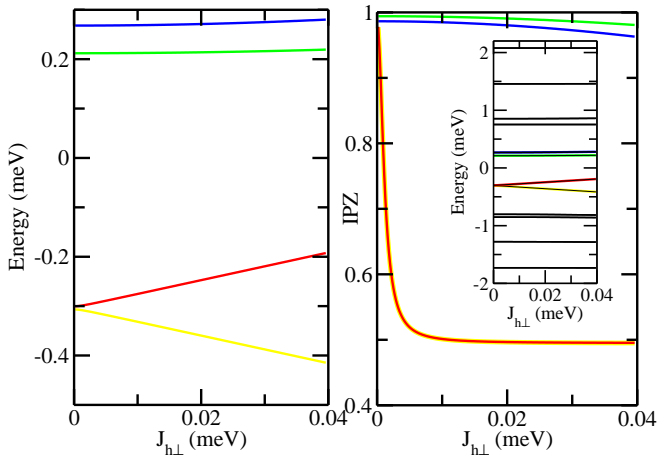}
\end{center}
\caption{ \label{fig2}(Color online) Left panel: Evolution
of the exciton levels as a function of the LH-HH mixing
parameter $J_{h\perp}$. Right panel: Evolution of the IPZ
as a function of the LH-HH mixing parameter. The inset
presents the evolution of the energy for all the 24 exciton
levels.}
\end{figure}

In figure (\ref{fig2})  we plot the evolution of both the
energy  (left panel)  and the $IPZ$  (right panel) as a
function of $J_{h\perp}$,  the LH-HH mixing parameter, of
four states denoted by their dominant component at
$J_{h\perp}=0$. For our choice of exchange constants, two
of them $|+1, -\frac{3}{2}\rangle$ and $|-2, -\frac{1}{2}\rangle$ are
almost degenerate at $J_{h\perp}=0$,  which means that the
hole-Mn exchange compensates the dark-bright splitting, and
couple these states via a hole-Mn spin flip.  As a result,
their energy levels split linearly as a function of
$J_{h\perp}$ and the wave-functions have a large weight on
the two product states for finite $J_{h\perp}$. In
contrast, the other two levels shown in figure
(\ref{fig2}), $|+1, -\frac{1}{2}\rangle$ and
$|-2,-\frac{3}{2}\rangle$ are
not coupled via hole-Mn spin flip. As a result, their
energies shift as a function of $J_{h\perp}$ due to
coupling to other states, and their IPZ undergoes a minor
change, reflecting moderate mixing.

\subsubsection{Exchange induced dark-bright mixing}
The most conspicuous experimentally observable consequence
of the exchange induced mixing, is    the transfer of
optical weight from the bright to the dark exciton, which
results in the observation of more than 6 peaks in the PL.
This can be understood as follows.  The spin-flip part of
the hole-Mn interaction couples  the bright exciton
$|+1,M_z\rangle$ to the dark exciton  $|-2,M_z+1\rangle$. Thus, a
state  with dominantly dark character  $|-2,M_z+1\rangle$ and energy given, to first order, by
that of the dark exciton, has a small but finite
probability of emitting a photon through its bright
component, via a Mn-hole coherent spin-flip. Thus, PL is
seen at transition energy of the dark exciton.  Reversely,
nominally bright excitons loose optical weight due to their
coupling to the dark sector.  Importantly,  the emission of
a photon from a dark exciton with dominant Mn spin
component  $M_z$, entails carrier-Mn spin exchange,  so
that the ground state has $M_z\pm 1$.

According to previous theory work  \cite{Fernandez-Rossier_PRB_2006} the  rate for the
emission of a circularly polarized photon   from the
exciton state $\Psi_n$ to the ground state $\phi_{m}$
reads:
\begin{equation}
\Gamma^{\pm}_{n,m} =\Gamma_0
\left|\sum_{M_z} \phi_m(M_z) \Psi^*_n(M_z,X=\pm1)\right|^2
\label{rate-photon}
\end{equation}
where
\begin{equation}
\Gamma_0\equiv \frac{3 \omega^3d_{cv}^2}{4 \pi \epsilon \hbar c^3}
\end{equation}
is the recombination rate of the bare exciton, $\omega$ is
the frequency associated to the energy difference between
the exciton state $n$ and the ground state $m$, $c$ is the
speed of light, $\epsilon$ is the dielectric constant of
the material, $d_{cv}$ is the dipole matrix element.  From
the experiments, we infer $\Gamma_0= $0.5 ns$^{-1}$

\begin{figure}[hbt]
\begin{center}
\includegraphics[angle=0,width=1\linewidth]{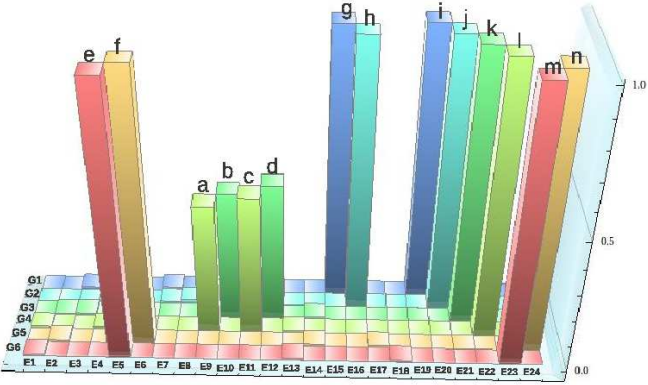}
\end{center}
\caption{ \label{fig2b}(Color online) Recombination rates
of the excitons levels in a Mn doped quantum dot
$\Gamma_i/\Gamma_0$: $a:|+2,+\frac{1}{2}> \rightarrow |+\frac{3}{2}>$,
$b:|-2,-\frac{1}{2}> \rightarrow |-\frac{3}{2}>$, $c: |-1,+\frac{3}{2}> \rightarrow
|+\frac{3}{2}> $, $d:|+1,-\frac{3}{2}> \rightarrow |-\frac{3}{2}>$, states from e
to n are bright excitons. Other states are mainly dark
excitons with a small bright component.}
\end{figure}

In the absence of spin-flip terms, the matrix
$\Gamma^{\pm}_{n,m}$ would have only non-zero elements for
$n=|X=\pm1 ,M_z\rangle$ states connected to $m=M_z$ states.
The presence of spin-flip terms in the Hamiltonian enables
the recombination from exciton states with dominant dark
component. In figure  (\ref{fig2b}) we represent the matrix
elements  $\Gamma^{\pm}_{n,m}/\Gamma_0$ for $j_{eh}=-0.73meV$,
$j_h=0.36meV$, $j_e=-0.09meV$, $j_{h\perp}=0.036meV$,
$D=0.01meV$ and $E=0meV$. It is apparent that the
recombination rates from the dark states are, at least, 2
times smaller (a and b) than those of the bright states.
For $\Gamma_0$= 0.4 ns$^{-1}$, the lifetime of the dark
excitons (a and b) are in the range of 3 ns. Thus, this
provides a quite efficient Mn spin relaxation mechanism,
provided that a dark exciton is present in the quantum dot.

The recombination rate matrix, together with the
non-equilibrium occupation of the exciton states, $P_n$,
determines the PL spectrum\cite{Fernandez-Rossier_PRB_2006}
for $\pm$ circular polarization:
\begin{equation}
{\cal I}_{\rm PL}^{\pm}(\omega)= \sum_{n,m,} P_n \Gamma^{\pm}_{n,m} \delta\left(\hbar\omega- E_n-E_m\right)
\end{equation}
In a typical  PL spectrum\cite{Le-Gall_PRB_2010}, the dark
peaks are, at most, 2 times smaller than the bright peaks.
Since  $\Gamma^{\pm}_{n,m}$ is at least 2 times smaller for
dark states, this implies a larger occupation of the dark
states. Thus, we can infer a transfer from the optical
ground state to the dark exciton states, via the bright
exciton states.  This transfer requires an incoherent spin
flip of either the electron or hole.  Below we show that
phonon induced hole-spin relaxation provides the most
efficient channel for this bright to dark conversion.

\section{Mn spin relaxation due to spin-phonon coupling}
\label{Mn-phonon}

In this section we discuss the Mn spin relaxation in the
absence of excitons.  In the absence of carriers and given
the fact that  Mn-Mn distance is comparable to the dot-dot
distance (100 nm for a dot and Mn density of about
$10^{10}cm^{-2}$), which makes direct super-exchange
negligible,  the Mn-phonon coupling should be the dominant,
albeit small, Mn spin relaxation mechanism.   Transverse
phonons induce local rotations of the lattice. Since the
crystal field, together with spin orbit coupling,
determines the Mn magnetocrystalline anisotropy, the phonon
induced lattice rotation \cite{Chudnovsky_PRB_2005} acts as
a stochastic torque on the Mn spin, resulting in spin
relaxation.

The atomic displacement at point $\vec{r}$ in the crystal
is expressed in terms of the phonon operators with wave
vector $\vec{q}$, polarization mode $\lambda=T1,T2,L$,
frequency $\omega_{\lambda}(\vec{q})$ and polarization
vector $\vec{e}_{\lambda}(\vec{q})$\cite{Cardona-Yu}:
\begin{equation}
\vec{u}(\vec{r})=\sum_{\vec{q},\lambda}
{\cal U}_{\lambda}(\vec{q})
\vec{e}_{\lambda}(\vec{q})
\left(b^{\dagger}_{\vec{q},\lambda}  +  b_{-\vec{q},\lambda}\right) e^{i\vec{q}\cdot\vec{r}}
\label{u-vs-phonon}
\end{equation}
where
\begin{equation}
{\cal U}_{\lambda}(\vec{q})=
 \sqrt{\frac{\hbar}{2 \rho\omega_{\lambda }(\vec{q})V}}
\end{equation}
and $V$ and $\rho$ are  the volume of the crystal and the
mass density respectively. In a zinc-blende structure there
are two  transverse acoustic (TA) phonon branches and one
longitudinal acoustic branch (LA). Following
Woods\cite{Woods04} we have:
\begin{equation}
\vec{e}_{TA1}= \frac{1}{q q_{\perp}} \left(q_x q_z, q_y q_z, -q_{\perp}^2\right)
\end{equation}
\begin{equation}
\vec{e}_{TA2}= \frac{1}{ q_{\perp}} \left(q_y ,-q_x,0\right)
\end{equation}
where $q\equiv |\vec{q}|$ and
$q_{\perp}=\sqrt{q_x^2+q_y^2}$. These vectors  satisfy
$\vec{q}\cdot\vec{e}_{TAi}=0$, and
$\vec{e}_{TAi}\cdot\vec{e}_{TAj}=\delta_{ij}$ The
longitudinal mode has $\vec{e}_{LA}= \frac{1}{q}\vec{q}$.

The lattice rotation vector is given by
\cite{Chudnovsky_PRB_2005}
 \begin{equation}
 \vec{\delta \Phi}(\vec{r})=\vec{\nabla}\times\vec{u}(\vec{r})
 \end{equation}
so that only the transverse modes contribute.  Within this
picture, the Mn spin-phonon coupling can be written
as\cite{Chudnovsky_PRB_2005}:
\begin{equation}
{\cal V}_{\rm M-ph}= i \left[{\cal H}_0,\vec{M}\right]\cdot \vec{\delta \Phi}(\vec{r}_{Mn})
\label{M-ph}
\end{equation}
Without loss of generality we can set the Mn position as
the origin, $\vec{r}_{Mn}=0$. Equation (\ref{M-ph}) couples
the Mn spin to a reservoir of phonons whose non-interacting
Hamiltonian is
 \begin{equation}
{\cal H}_{\rm ph}=\sum_{\vec{q},\lambda} \hbar \omega_{\lambda }(\vec{q})
b^{\dagger}_{\vec{q},\lambda}   b_{\vec{q},\lambda}
\label{free-phonon}
\end{equation}

Within the standard system plus reservoir master equation
approach, we have derived the scattering rate from a state
$n$ to a state $n'$, both eigenstates of the single Mn
Hamiltonian ${\cal H}_0$, due the emission of a phonon. In
order to use  a general result  for that rate
(\ref{general-rate}), derived in the appendix
(\ref{phonon-app}), we need to express the spin-phonon
coupling  (\ref{M-ph}) using the same notation than in
equation (\ref{general-coupling}):
\begin{equation}
{\cal V}^{n,n'}_{\vec{q},\lambda}=i^2
{\cal U}_{\lambda}(\vec{q})
 \vec{f}_{n,n'}
   \cdot \left(\vec{q}\times \vec{e}_{\lambda}(\vec{q})\right)
\label{M-ph2}
\end{equation}
where
\begin{equation}
\vec{f}_{n,n'}\equiv\langle n|\left[{\cal H}_0,\vec{M}\right]|n'\rangle
\end{equation}

We compute now the scattering rate due to a single phonon
emission assuming 3 dimensional phonons described  above.
The rate reads:
\begin{equation}
\Gamma_{n\to n'}=\frac{|\Delta|^3}{12\pi\rho\hbar^4
c^5}(n_B(\Delta)+1) \sum_{b,b'=x,y,z} f^b_{n',n} (f^{b'}_{n',n} )^{\ast}
\label{rate-Mn}
\end{equation}
where $c= 1.79\textrm{kms}^{-1}$ is the CdTe speed of
sound\cite{merle84}, $\rho=5870\text{kgm}^{-3}$ is the mass density of
the CdTe unit cell\cite{collins80} and $n_B(\Delta)\equiv
\frac{1}{e^{\beta|\Delta|}-1}$. The  $|\Delta|^3$ factor
comes from the dependence of the phonon density of states
on the energy.

\subsection{Mn spin relaxation in the optical ground state}

We now discuss  the relaxation of the Mn electronic spin
due to spin-phonon coupling without an exciton in the
quantum dot. According to our experimental
results\cite{Le-Gall_PRL_2009,Le-Gall_PRB_2010}, the Mn
spin relaxation time in our samples is at least 5$\mu$s.

\begin{figure}[t]
\begin{center}
\includegraphics[angle=0,width=1\linewidth]{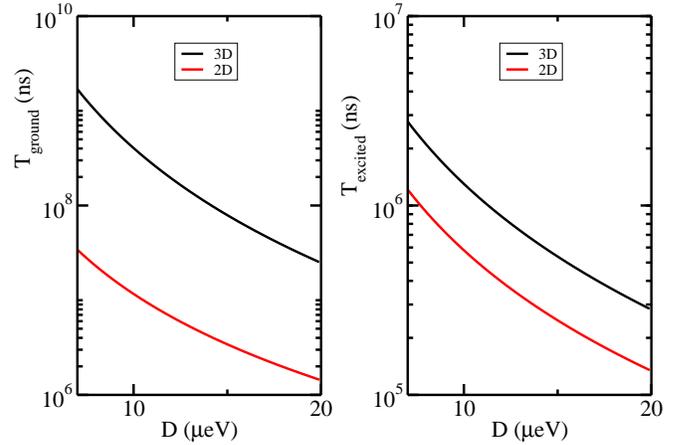}
\end{center}
\caption{ \label{fig3}(Color online) Left panel: Lifetime
of the (+$\frac{5}{2}$ to +$\frac{3}{2}$) transition in the optical ground
state at zero field as a function of the magnetic
anisotropy energy splitting $D$. Right panel: Lifetime of
the same transition in the presence of a $+1$ exciton for
different values of $D$. The rates are calculated for a 3
dimensional (3D) and a 2 dimensional (2D) density of states
of acoustic phonons.}
\end{figure}

If we take $E=0$, the transition rate between the excited
states $| \phi_n\rangle=|M_z=+\frac{5}{2}\rangle$ and
$| \phi_{n'}\rangle=|M_z=+\frac{3}{2}\rangle$, via a phonon
emission, is given by:
\begin{equation}
\Gamma_{n\to n'}=\frac{640|D|^5}{3\pi\rho\hbar^4c^5}(n_B(\Delta)+1)
\end{equation}
The dependence on $D^5$ comes both  from the density of
states of phonons $\rho\propto \omega^3$ and the square of
the Mn phonon coupling, which is proportional to the
anisotropy, and gives the additional  $D^2$ factor. Whereas
the uniaxial anisotropy of Mn in CdTe quantum wells has
been determined by EPR  \cite{Qazzaz}, the actual value for
Mn in quantum dots is not known and can not be measured
directly from single exciton spectroscopy of neutral dots.
Therefore, in figure (\ref{fig3}) we plot the lifetime for
the transition of the Mn spin from $\frac{5}{2}$ to $\frac{3}{2}$, due to a
phonon emission, as a function of $D$. We take $D$ in a
range around the value reported for CdTe:Mn epilayers,
$D=12\mu$eV \cite{Qazzaz}. We find that the spin lifetime
of Mn in the optical ground state can be very large. Even
for $D=20\mu eV$ the Mn spin lifetime is in the range of
0.1 seconds, well above the lower limit for the Mn spin
relaxation reported
experimentally\cite{Le-Gall_PRL_2009,Le-Gall_PRB_2010}.
Whereas we can not rule out completely that the Mn spin
lifetimes that long, there are other spin relaxation
mechanisms that might be more efficient that  the Mn-phonon
coupling considered above, like the coupling of the Mn
electronic spin to nuclear spins of Mn and the host atoms\cite{Le-Gall_PRL_2009}.


Since part of the  $\Delta^5$ scaling arises from the
$\omega^3$ scaling of the phonon density of states, we have
explored the possibility that phonons localized in the
wetting layer could be more efficient  in relaxing the Mn
spin. For that matter we have considered a toy model of two
dimensional phonons confined in a slab of thickness
$L=2\,nm$. The resulting Mn spin relaxation rate for those
reads:
\begin{equation}
\Gamma_{n\to n'}= \frac{\Delta^2}{16\hbar^3c^4\rho W}(n_B(
\Delta)+1) \sum_{b,b'=x,y,z}A_{b,b'} f^b_{n',n} (f^{b'}_{n',n} )^{\ast}
\label{rate-Mn2D}
\end{equation}
where W is the width of the sample and A is a diagonal
matrix with $A_{xx}=1$, $A_{yy}=1$, $A_{zz}=2$.

In figure (\ref{fig3}) we plot the associated spin lifetime
in this case, taking $W=2nm$ and show how it is at least 100 shorter than
for 3D phonons, but still we would have $T_1\simeq1$ ms for
$D=20\mu$eV.

\subsection{Mn spin relaxation in the presence of an exciton }
Here we discuss how the Mn spin relaxation due to Mn-phonon
coupling is modified when an exciton is interacting with
the Mn. The Mn-phonon coupling is still given by
Hamiltonian (\ref{M-ph}), with ${\cal H}_0$ given by eq.
(\ref{H00}).   We assume that the only effect of the
exciton on the Mn is to change the energy spectrum and mix
the spin wave-functions, giving rise to larger spin
relaxation rates, due to the larger exchange-induced energy
splittings.

In the presence of the exciton, the Mn-phonon coupling
results in transitions between different exciton-Mn  spin
states, $n$ and $n'$.  As we did in the case of the Mn
without excitons,  we need to express the spin-phonon
coupling  (\ref{M-ph}) using the same notation than in
equation (\ref{general-coupling}).

For that matter we define the matrix elements
\begin{eqnarray}
\vec{F}_{n,n'}\equiv\langle \Psi_n|\left[{\cal H}_0,\vec{M}\right]|\Psi_n'\rangle=\nonumber\\
\sum_{X,M_z,M_z'} \Psi_n(X,M)^* \Psi_n(X,M')^*\vec{f}_{M,M'}
\end{eqnarray}
where
 \begin{equation}
\vec{f}_{M,M'}\equiv\langle M|\left[{\cal H}_0,\vec{M}\right]|M'\rangle
\end{equation}
$M$ and $M'$ stand for eigenstates of the Mn spin operator
$M_z$. Thus, in the exciton-Mn spin states basis, the
Mn-phonon coupling reads:
\begin{equation}
{\cal V}^{n,n'}_{\vec{q},\lambda}=i^2
{\cal U}_{\lambda}(\vec{q})
 \vec{F}_{n,n'}(M)
   \cdot \left(\vec{q}\times \vec{e}_{\lambda}(\vec{q})\right)
\label{M-ph3}
\end{equation}
Notice how if we neglect the spin mixing of the exciton
states we have $\vec{F}_{n,n'}= \vec{f}_{M,M'}$ and the
only difference in the scattering rates arises from the
larger  energy splittings in the presence of the exciton.

Using the  equation (\ref{general-rate}) for the phonon
induced spin relaxation rate, and in analogy with equation
(\ref{rate-Mn}) we write:
\begin{equation}
\Gamma_{n\to n'}=\frac{|\Delta|^3}{12\pi\rho\hbar^4
c^5}(n_B(\Delta)+1) \sum_{b,b'=x,y,z} F^b_{n,n'} (F^{b'}_{n,n'} )^{\ast}
\label{rate-X-Mn-phonon}
\end{equation}

In figure (\ref{fig3}) we see how Mn-phonon spin relaxation
is much faster in the presence of the exciton. Ignoring the
difference arising from the spin mixing, we can write the
ratio of the rates as:
\begin{equation}
\frac{\Gamma_{n\to n'}(X)}{\Gamma_{n\to n'}(G)}= \left(\frac{\Delta_X}{\Delta_G}\right)^3
\end{equation}
The energy splitting associated to the $\frac{5}{2}$ to $\frac{3}{2}$ spin
flip in the ground state is $4D$.  In the presence of the
exciton the energy splitting of the same transition would
be $4D+ j_h -j_e$. If we take $D=12\mu$eV, $j_h=360\mu$eV
and $j_e=-90\mu$eV the ratio yields $\approx 10^3$. From
the experimental side we know that $T_{1G}> 5\mu s$  and,
in the presence of the exciton $T_1\simeq 50 ns$. Thus, the
ratio  could be accounted for by this mechanism. However,
in order to have $T_{1G}=5\mu $s we would need to assume an
unrealistically large value for $D$.  Thus, we think that
another spin relaxation mechanism must be operative in the
system when the exciton is in the dot which makes it
possible to control the spin of the Mn in a time scale of
50 ns.  In the next sections we discuss the hole spin
relaxation due to phonons as the mechanism that, combined
with Mn-carrier exchange, yields a quick Mn spin relaxation
in the presence of the exciton.

\section{Hole spin relaxation in non magnetic dots}
\label{hole-spin}
\subsection{Hole-phonon coupling}

We now consider the relaxation of the hole spin due to
hole-phonon coupling. We consider first the case of undoped
quantum dots. The  coupling of the spin of the hole to
phonons can be understood extending the Bir Pikus
Hamiltonian to the case of inhomogeneous strain associated
to lattice vibrations:
\begin{equation}
\epsilon_{ij}(\vec{r})\equiv\frac{1}{2}\left(\frac{\partial u_i}{\partial r_j} + \frac{\partial u_j}{\partial r_i}\right)
\label{strain-tensor}
\end{equation}
It is convenient to write the strain tensor field as:
\begin{equation}
\epsilon_{ij}(\vec{r}) =\sum_{\vec{q}} e^{i\vec{q}\cdot\vec{r}}\epsilon_{ij}(\vec{q})
\end{equation}
so that we write:
\begin{equation}
\epsilon_{ij}(\vec{q})= \frac{1}{2}\sum_{\lambda}
{\cal U}_{\lambda}(\vec{q})\left(b^{\dagger}_{\vec{q},\lambda}  +  b_{-\vec{q},\lambda}\right)
 \left(q_j e^{i}_{\lambda}(\vec{q})+ q_i e^{j}_{\lambda}(\vec{q})\right)
\end{equation}

\begin{figure}[t]
\begin{center}
\includegraphics[angle=0,width=1\linewidth]{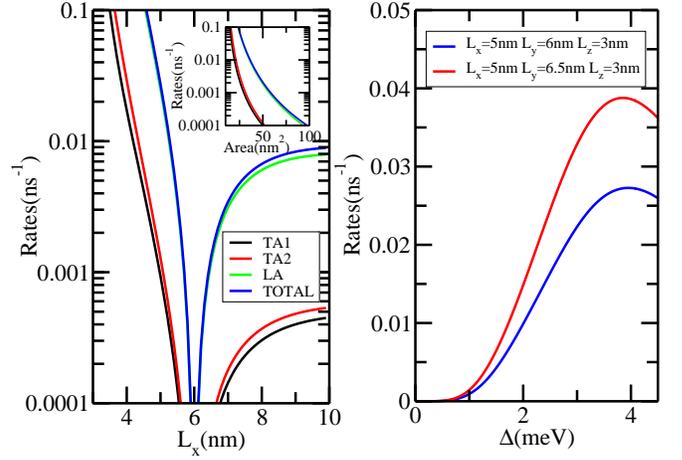}
\end{center}
\caption{ \label{fig4}(Color online) Left panel: Hole
spin-flip rate as a function of the dot size and shape.
$L_y$ is fixed at 6nm, scaning of $L_x$ change the LH-HH
mixing. In the inset the ratio $L_y/L_x$ is fixed at 1.2
so the shape of the dot are fixed, only the size of the dot
changes. One can see that the hole spin-flip rate is a size
sensitive quantum quantity, the rate is a semi-exponential
function of the size of the dot. Right panel: Hole
spin-flip rate as a function of the energy splitting for
two different values of the quantum dot anisotropy, ie
LH-HH mixing.}
\end{figure}

We consider the coupling of the ground state doublet,
formed by states $\Uparrow$ and $\Downarrow$, to the phonon
reservoir \cite{Roszak_PRB_2007}. The effective hole-phonon
Hamiltonian is obtained by projecting the BP Hamiltonian
onto this subspace:
\begin{equation}
{\cal V}_{\rm h-phonon}= \sum_{ij, \vec{q}\sigma_h,\sigma_h'}  I^{\sigma_h,\sigma_h'}_{ij}(\vec{q})
|\sigma_h\rangle\langle\sigma_h'|
 \epsilon_{ij}(\vec{q})
\label{hole-phonon1}
\end{equation}
Here $|\sigma_h\rangle$ denotes the  quantum dot state
defined in eq. (\ref{WF-total}) and the coupling constant
reads
\begin{eqnarray}
I^{\sigma_h,\sigma_h'}_{ij}(\vec{q})\equiv
\sum_{j_z,j_z'} V_{ij} {\cal C}_{h}^*(j_z){\cal
C}_{h'}(j_z')\langle j_z|J_i J_j |j_z'\rangle I_{\vec
q}\end{eqnarray} where $I_{\vec q}= \int
|\psi_0(\vec{r})|^2 e^{i\vec{q}\cdot\vec{r}}d\vec{r}$.
Hamiltonian  (\ref{hole-phonon1}) shows how  the
absorption or emission of a phonon can induce a transition
between  the two quantum dot hole states, $\Uparrow$ and
$\Downarrow$.

We now calculate the time scale for the spin relaxation of
a single hole in a non magnetic dot under the influence of
an applied magnetic field so that the hole ground state
doublet is split in energy. In order to compute  the
transition rate for decay of the hole from the excited to
the ground state we use again the general equation
(\ref{general-rate}). For that matter, we express the
hole-spin coupling (\ref{hole-phonon1}) as:

\begin{equation}
{\cal V}_{\rm h-phonon}= \sum_{ \vec{q},\lambda\sigma_h,\sigma_h'}
{\cal V}^{\sigma_h,\sigma_h'}_{\vec{q},\lambda}
|\sigma_h\rangle\langle\sigma_h'|
\left(b_{\lambda q}^{\dagger}+b_{\lambda,-q}\right)
\label{hole-phonon2b}
\end{equation}
where
\begin{equation}
{\cal V}^{\sigma_h,\sigma_h'}_{\vec{q},\lambda} =\frac{i}{2} \sum_{i,j}
I^{\sigma_h,\sigma_h'}_{ij}(\vec{q})
{\cal U}_{\lambda}(\vec{q})
   \left(q_i e^j_{\lambda}(\vec{q}) +
   q_j e^i_{\lambda}(\vec{q}) \right)
\label{hole-phonon2}
\end{equation}

\subsection{Calculation of hole spin-flip rates with simple model}
In order to illustrate the physics of the phonon-driven
hole spin relaxation we consider the case of a single hole
in a non-magnetic dot under the influence of an applied
magnetic field. For that matter, we compute the Hamiltonian
(\ref{hole-phonon2}) using the wave functions from the
simple model of confined holes defined in eq.
(\ref{WF-KL-nostrain}).  We focus on the non-diagonal terms
in the hole spin index, i.e., the terms that result in
scattering form $\Uparrow$ to $\Downarrow$ due to phonon
emission.

Importantly, the BP Hamiltonian couples hole states that
differ in, at most, two units of $J_z$. Thus, in the
absence of LH-HH mixing, the BP Hamiltonian does not couple
directly the $\Uparrow$ and $\Downarrow$ states.
Transitions between $\Uparrow$ and $\Downarrow$ states, as
defined in eq. (\ref{WF-total}),  are only possible,
through  one  phonon processes,  through the
$\epsilon_{yz}(J_yJ_z+J_zJ_y)$ and
$\epsilon_{zx}(J_zJ_x+J_xJ_z)$  terms in the Hamiltonian.
After a straightforward calculation we obtain:


 \begin{equation}
{\cal V}^{\Uparrow,\Downarrow}_{\vec{q},\lambda} =\frac{i}{2}
\sqrt{3}d Sin\left(\frac{\theta_1-\theta_2}{2}\right)
\left( \epsilon_{yz}(\vec{q})-i\epsilon_{zx}(\vec{q})\right)
\label{hole-phonon3}
\end{equation}
The important role played by the $LH-HH$ mixing
$\theta_{1,2}$ is apparent. Using equation
(\ref{general-rate}) it is quite straightforward to compute
the rate for the 3 phonon branches. They are all
proportional to
\begin{equation}
\Gamma^{0}_{\Uparrow\rightarrow \Downarrow } = \frac{1}{18\pi} D_{u'}^2 Sin^2\left(\frac{\theta_1-\theta_2}{2}\right)
   \frac{\Delta^3 }{\rho\hbar^4  c^5}
   \label{Non-magnetic-hole-rates}
\end{equation}
with coefficients $\frac{7}{5}$, $1$ and $\frac{8}{5}$ for
the TA1, TA2 and L modes respectively. Here, $D_{u'}$
stands for the deformation potential of
Kleiner-Roth\cite{kleiner}, following reference
\onlinecite{willatzen}, $D_{u'}=-\frac{3\sqrt{3}d}{2}$.
$\rho$ for the mass density of CdTe,  $c$ for its
transverse speed of sound,  and $\Delta$ for the energy
splitting between the $\Uparrow$ and $\Downarrow$ states,
which is proportional to the external magnetic field $B$.
In figure (\ref{fig4}) we plot the  rates
$\Gamma_{TA1},\Gamma_{TA2},\Gamma_{L}$, as well as their
sum as function of the dot size (left panel) and as a
function of the energy splitting between the initial and
final hole state,  $\Delta$ (right panel).  We see how hole
spin relaxation rates can be in the range of $\Gamma \simeq
1/(40 ns)$.

The results of figure (\ref{fig4}) suggest that for
sufficiently high $\Delta$, as those provided by the
Mn-hole exchange, the hole spin can relax in a time scale
of 30$ns$.  These numbers are in the same range than those
obtained by Woods et al \cite{Woods04}. As we discuss in
the next section, these spin flips, together with
Mn-carrier exchange, can also induce Mn spin relaxation in
a time scale much shorter than the one due to Mn-phonon
coupling only.

Importantly, the rate is finite only if
$\theta_1-\theta_2\neq0$, which is the case in the presence
of an applied magnetic field.  This indicates that,  within
the simple model of eq. (\ref{WF-KL-nostrain}),  the
non-diagonal terms in the hole-phonon Hamiltonian
(\ref{hole-phonon1}) vanishes identically. This is not a
general feature of (\ref{hole-phonon1}), but rather an
particular property of the simple model
(\ref{WF-KL-nostrain}). In particular,  the non-diagonal
term in (\ref{WF-KL-nostrain}) is non-zero at zero field as
soon as the ($+\frac{3}{2},-\frac{1}{2}$) states have also some weight on
the $+\frac{1}{2}$ components, which happens both when a more
realistic model for confinement is used or when homogeneous
strain components $e_{yz}$ and $e_{zx}$ are included.

\section{ Spin relaxation in magnetic dots due to hole-phonon coupling}
\label{hole-phonon-magnetic}

The results of the previous sections indicate that, because
of their coupling to  phonons,  the hole spin lifetime in a
non-magnetic dot is much shorter than the Mn spin lifetime.
Here we explore the consequences of this phonon-driven hole
spin relaxation for the single exciton states in a dot
doped with one magnetic atom. The leading process results
in a Mn spin conserving  decay from the bright  exciton  to
the dark exciton state,  via hole-spin flip in  a time
scale in the 10 ns range.  Combined with the optical
recombination of the dark state, made possible via Mn-hole
or Mn-electron spin flip,  provide a pathway for exciton
induced Mn spin relaxation in a time scale under 100 ns, as
observed experimentally
\cite{Le-Gall_PRL_2009,Goryca_PRL_2009,Le-Gall_PRB_2010}.

\begin{figure}[t]
\begin{center}
\includegraphics[angle=0,width=0.7\linewidth]{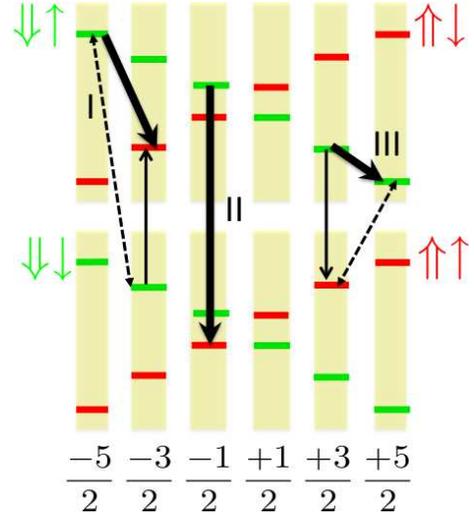}
\end{center}
\caption{ \label{fig5}(Color online) Scheme of the Mn spin
flip channels due to the combined action of hole-phonon
coupling and carrier-Mn exchange.}
\end{figure}

We also explore the  scattering between two bright states
enabled by the combination of phonon induced hole spin
relaxation and Mn-carrier exchange. The lifetimes of these
processes is in the range of $10^3$ ns and higher, and
therefore they are probably not determinant for the optical
orientation of the Mn spin in the sub-microsecond scale.

\subsection{Exciton-phonon coupling in magnetic dots}

The Hamiltonian that couples the exciton states $\Psi_n$ to
the phonons is derived by projecting the hole-phonon
coupling Hamiltonian (\ref{hole-phonon1}) onto  the exciton
states  (\ref{exciton-vs-P}) .  The result reads:
\begin{equation}
{\cal V}_{\rm X-phon}= \sum_{n,n' \vec{q}\lambda}
|\Psi_n\rangle\langle\Psi_{n'}|
{\cal V}^{n,n'}_{\vec{q},\lambda}
\left(b_{\lambda q}^{\dagger}+b_{\lambda,-q}\right)
\label{X-phonon1}
\end{equation}
where
\begin{equation}
{\cal V}^{n,n'}_{\vec{q},\lambda}=\sum_{M_z,\sigma_e,\sigma_h,\sigma_h'}
 {\cal V}^{\sigma_h,\sigma_h'}_{\vec{q},\lambda} \Psi_n(M_z,X)
 \Psi^*_{n'}(M_z,X')
 \label{X-phonon2}
\end{equation}
where $X=(\sigma_e,\sigma_h)$ and $X'=(\sigma_e,\sigma_h')$ (same electron spin) and ${\cal V}^{\sigma_h,\sigma_h'}_{\vec{q},\lambda}$ is given by equation (\ref{hole-phonon2}).

\subsection{Qualitative description of the spin relaxation processes}

In order to describe qualitatively the variety of different
processes accounted for by Hamiltonian (\ref{X-phonon1}) it
is convenient to consider an initial state $\psi_n$ as a
linear combination of  a dominant component $|M\rangle
|\downarrow\rangle_e |\Uparrow\rangle_h$ plus  a minor
contribution of  two  dark components, which arise from the
coherent exchange of the Mn with either the electron or the
hole:
\begin{eqnarray}
|\psi_n\rangle &=&  |M\rangle |\downarrow\rangle_e |\Uparrow\rangle_h +\nonumber\\
+&&\epsilon_e  |M-1\rangle |\uparrow\rangle_e |\Uparrow\rangle_h +
\epsilon_h  |M+1\rangle |\downarrow\rangle_e |\Downarrow\rangle_h
\label{initial}
\end{eqnarray}
where $\epsilon_e\propto j_e/j_{eh}$ and $\epsilon_h\propto
 j_h/j_{eh} $ are small dimensionless coefficients
that can be obtained doing perturbation theory.

\begin{figure}[t]
\begin{center}
\includegraphics[angle=0,width=1\linewidth]{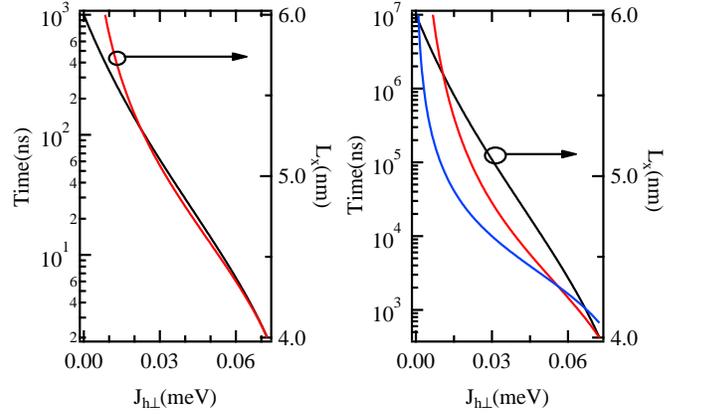}
\end{center}
\caption{ \label{fig6}(Color online) Calculated rates for
the transitions between exciton states in a Mn-doped
quantum dot due to hole-phonon coupling. Left panel, red
line, transition from $|-1, -\frac{5}{2}>$ to $|+2,
-\frac{5}{2}>$, Right panel, red line, transition from$|-1,
-\frac{5}{2}>$ to $|-2, -\frac{1}{2}>$.  blue
line, transition from $|-1, -\frac{5}{2}>$ to $|+1,
-\frac{3}{2}>$. L$_y$ is fixed at 6nm and scanning L$_x$
changes the LH-HH mixing parameter J$_{h\perp}$.}
\end{figure}

Depending on the elementary process that takes place, there
are several possible final states:
\begin{enumerate}
\item Hole spin relaxation. In this case the final state would be dominantly a dark exciton whose
wave function read:
\begin{eqnarray}
|\psi_{n'}\rangle &=&  |M\rangle |\downarrow\rangle_e |\Downarrow\rangle_h + O(\epsilon)
\end{eqnarray}
and the scattering rate $\Gamma_0$ would be proportional to $|I^{\Uparrow,\Downarrow}|^2$.
This is process  II in figure (\ref{fig5}).

\item  Hole spin relaxation plus coherent hole-Mn spin flip.
This is process  III in figure (\ref{fig5})
 This can be realized through 2  dominant
channels. An incoherent hole spin flip will couple the
dominant component of the initial state, $|M\rangle
|~\downarrow\rangle_e |\Uparrow\rangle_h$ with a secondary
component $|M\rangle |\downarrow\rangle_e
|\Downarrow\rangle_h$ of the final state
\begin{eqnarray}
|\psi_{n'}\rangle &=&  |M-1\rangle |\downarrow\rangle_e |\Uparrow\rangle_h +\nonumber\\
+&&\epsilon_h  |M\rangle |\downarrow\rangle_e |\Downarrow\rangle_h  +O(\epsilon_e  )
\end{eqnarray}
In this case the final state is a bright exciton in the
same branch $+1$ than the initial state but the Mn
component goes from $M$ to $M-1$.

The second channel comes from the hole spin flip of the minority dark component of the initial state,
$ \epsilon_h  |M+1\rangle |\downarrow\rangle_e |\Downarrow\rangle_h $ which decays into the majority
component of the final state
\begin{eqnarray}
|\psi_{n'}\rangle &=&  |M+1\rangle |\downarrow\rangle_e |\Uparrow\rangle_h +O(\epsilon_{e,h}  )
\end{eqnarray}
Thus, in this second case a hole spin flips due to phonons,
plus a coherent Mn-electron spin flip connect the $X=+1, M$
initial state to the $X=-1, M+1$ state.  Thus, both the
initial and final state in this process are the same than
in the first channel, the rates for each would be
proportional to $\epsilon_h^2 \Gamma_0$, but the decay
pathways are different, and interferences are expected.

\item  Hole spin relaxation plus coherent electron-Mn spin flip.
This is process  I in figure (\ref{fig5})
As in the previous case, there are two channels for this
type of process. In the first channel, the majority
component of the initial state decays into a final state
given by:
\begin{eqnarray}
|\psi_{n'}\rangle &=&  |M-1\rangle |\uparrow\rangle_e |\Downarrow\rangle_h +\epsilon_{e}'
 |M\rangle |\downarrow\rangle_e |\Downarrow\rangle_h
 \label{final-elec1}
\end{eqnarray}
The incoherent hole spin flip connects the initial state
(\ref{initial}) to the final state (\ref{final-elec1})
through the minority component $ |M\rangle
|\downarrow\rangle_e |\Downarrow\rangle_h$ of the later.

The second channel comes from the hole spin flip of the minority dark component of the initial state,
$ \epsilon_e  |M-1\rangle |\uparrow\rangle_e |\Uparrow\rangle_h $ which decays into the majority
component of the final state
\begin{eqnarray}
|\psi_{n'}\rangle &=&  |M-1\rangle |\uparrow\rangle_e |\Downarrow\rangle_h +O(\epsilon_{e,h}  )
\end{eqnarray}
Thus, a hole spin flips due  the phonon, plus a coherent
Mn-electron spin flip connect the $X=+1, M$ initial state
to the $X=-1, M-1$ state. The scattering rate of these two
process scales as  $\epsilon_e^2\Gamma_0$
\end{enumerate}


\subsection{Calculation of the relaxation rates}

In order to implement equations
(\ref{X-phonon1},\ref{X-phonon2}) to compute scattering
rates, we use  the single particle basis for the holes done
with equations (\ref{WF-KL-nostrain}) which leads, at
finite magnetic field to the matrix element
(\ref{hole-phonon3}) that would be incorporated into
equations (\ref{X-phonon2}) to compute the rates using
equation (\ref{general-rate}).  As discussed above, a zero
field  model (\ref{WF-KL-nostrain}) yields a zero spin-flip
matrix element in equation (\ref{hole-phonon3}). This is a
feature of the simple hole model rather than an intrinsic
property of the system. Thus, for the sake of simplicity,
we compute the rates between exciton states by computing
the matrix element (\ref{hole-phonon3}) as if there was a
magnetic field that yields the energy splitting between the
initial and final exciton states equal to the splitting
produced by the exchange interaction with the Mn spin.

\begin{figure}[t]
\begin{center}
\includegraphics[angle=0,width=0.9\linewidth]{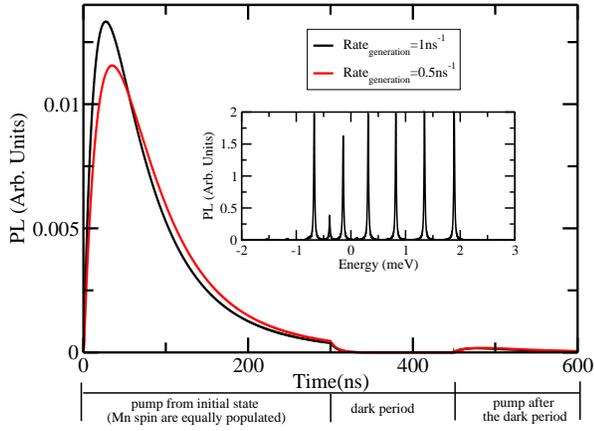}
\end{center}
\caption{\label{fig7}(Color online) Simulation of PL
intensity from state ($X=+1$,$M_z=-\frac{5}{2}$) under the
influence of a driving laser pumping the system resonantly
from optical ground state $M_z=-\frac{5}{2}$ to the excited state
($X=-1$, $M_z=-\frac{5}{2}$) for two laser intensities. The inset
is the PL spectrum assuming all the states are equally
populated. In the calculation, the quantum dot anisotropy
(L$_x$=5nm and L$_y$=6nm) controls the LH-HH mixing. The
other parameters are discussed in the text.}
\end{figure}

In the calculation of the rates we perform an additional
approximation: we only consider spin-flip terms in equation
(\ref{X-phonon2}) and we do exclude spin-conserving terms.
The results for transition rates from the state $n$ with
dominant ($-1,-\frac{5}{2}$)  to 3 possible final states with
dominant components ($+2,-\frac{5}{2}$), ($-2,-\frac{1}{2}$)
and ($+1,-\frac{3}{2}$) as a function of the spin-flip Mn
hole exchange $J_{h\perp}$, are shown in figure
(\ref{fig6}). The transition to the ($+2,-\frac{5}{2}$),
which only involves the irreversible spin flip of the hole
via a phonon emission  is the dominant process and has a
lifetime of 30 ns. The transition to the ($+1,-\frac{3}{2}$)
state requires both the hole spin flip and the Mn-hole spin
flip and it is 3 orders of magnitude less efficient.

Thus, these calculations indicate that the most likely
mechanism for Mn spin orientation in the presence of an
exciton combines a rapid bright-to dark conversion,
produced by phonon induced hole spin flip, and a dark to
ground transition, enabled by Mn-carrier spin exchange and
radiative recombination.

\begin{figure}[t]
\begin{center}
\includegraphics[angle=0,width=1\linewidth]{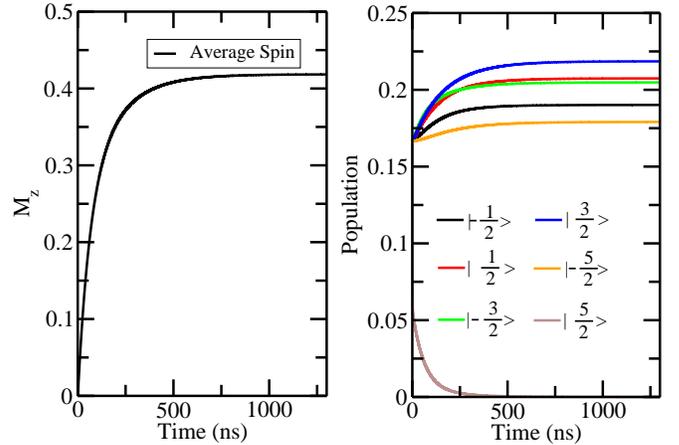}
\end{center}
\caption{\label{fig9}(Color online) Left  panel: average
magnetization and right panel: occupation of the different
spin states under optical pumping of the state ($X=-1$,
$M_z=-\frac{5}{2}$). Parameters are the same as for the calculation
presented in figure (\ref{fig7}).}
\end{figure}

\section{Laser driven spin dynamics }
\label{dynamics}
\subsection{Summary of scattering mechanisms and master equation}

The spin dynamics of a single Mn atom in a laser driven
quantum dot is described in terms of the 24 exciton states
$\Psi_n$ and the 6 ground states $\phi_m$. In the previous
sections we have calculated the scattering rates of these
states. They can be summarized as follows:
\begin{enumerate}

\item Transitions from the $\Psi_n$ to the $\phi_m$, via photon emission (eq. (\ref{rate-photon})).
In the case of bright excitons, this process is the
quickest of all, with a typical lifetime of 0.3 ns. In the
case of dark excitons the lifetime depends on the
bright/dark mixing, which is both level and dot dependent.
Dark lifetime ranges from twice the one of bright excitons
to 1000 times larger, ie, between 1 and 300 nanoseconds. In
any event, dark recombination  involves a Mn spin flip.

\item Transitions between different $\phi_m$ states, due to Mn spin phonon coupling (eq. (\ref{rate-Mn})).
The lifetimes of these transitions are, at least, 1ms (see right panel of figure \ref{fig3}).

\item Transitions between different exciton states $\Psi_n$ that flip the spin of the Mn only,
due to Mn-phonon coupling (eq. (\ref{rate-X-Mn-phonon}).
The lifetimes of these transitions are, at least, 0.1ms
(see left panel figure \ref{fig3}).

\item Transitions between exciton states due to hole-phonon coupling (eq. (\ref{X-phonon1}).
The bright to dark transition is the quickest  process with
a lifetime of about 30ns (see figure \ref{fig6}). Bright to
bright transitions, combining hole-phonon and Mn-carrier
interactions, have lifetimes in the 10 $\mu s$ range.

\end{enumerate}

In addition to these dissipative scattering processes, we
have to consider driving effect of the laser field,
described in the semiclassical approximation.  All things
considered, we arrive to  a master equation that describes
the evolution of the occupations $p_N$, where $N=(n,m)$
includes states both with and without exciton in the dot.
The master equation reads:
\begin{equation}
\frac{dp_N}{dt} = \sum_{N'} \Gamma_{N'\rightarrow N} p_{N'} -
\sum_{N'} \Gamma_{N\rightarrow N'} p_{N}
\label{master}
\end{equation}
Eq. (\ref{master}) is a system of 36 coupled differential
equations that we solve by numerical iteration, starting
from a thermal distribution for the  initial occupation
$p_N$. Since the temperature is larger than the energy
splitting in the ground state, but much smaller than the
band gap, at $t=0$ we have the six ground states with
similar occupation $P_m\simeq1/6$, $P_n\simeq0$. As a
result, the average magnetization, defined as:
\begin{equation}
\langle M_z \rangle = \sum_m p_m \langle \phi_m |M_z|\phi_m \rangle
\end{equation}
is zero, at zero magnetic field, as expected.

\begin{figure}[t]
\begin{center}
\includegraphics[angle=0,width=1\linewidth]{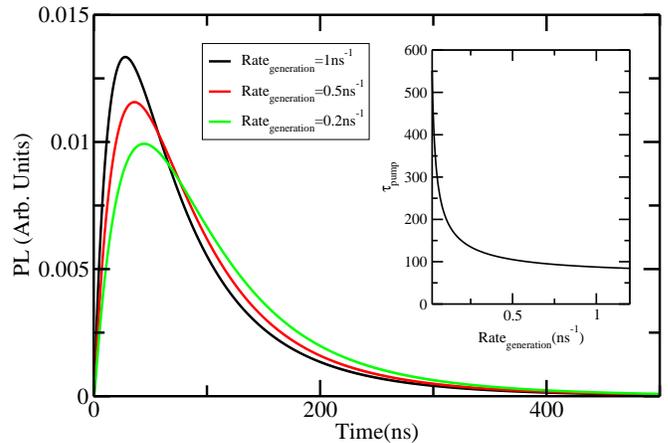}
\end{center}
\caption{ \label{fig10}(Color online) Evolution of Mn spin
orientation efficiency as a function of the laser power.
The pumping is detected in the PL intensity from state
($X=+1$,$M_z=-\frac{5}{2}$) under the influence of a driving laser
pumping the system from optical ground state $M_z=-\frac{5}{2}$ to
the excited state ($X=-1$, $M_z=-\frac{5}{2}$). The inset presents
the laser power dependence of $\tau_{pump}$, from where we
can see that the efficiency of the pumping gets higher with
the increasing of the laser power.}
\end{figure}

\subsection{Optical Mn spin  orientation }

Under the action of the laser, the exciton states become
populated and, under the adequate pumping conditions,  the
average Mn magnetization $\langle M_z \rangle$  acquires a
non-zero value. This transfer of angular momentum, known as
optical Mn spin orientation  has been observed
experimentally \cite{Le-Gall_PRL_2009} and predicted
theoretically\cite{Govorov05}.  It results from a decrease
of the Mn spin lifetime in the presence of the exciton in
the dot.  In that circumstance, the laser transfer
population from the $M_z$ state to the $X,M_z$ state. The
enhanced relaxation transfers population from $X,M_z$ to
$X,M_z'$ and the recombination to  $M_z'$ state. Thus, if
the laser is resonant with a single $M_z$ to $X,M_z$
transition, the $M_z$ state is  depleted, which results in
a decrease of the PL coming both from the $X,M_z$ and the
$-X,M_z$ transitions.

\begin{table}
\begin{center}
\begin{tabular}{ccc}
Quantity & Symbol & Value  \\
\hline \hline
Hole-Mn exchange & $j_h$ &  0.31 meV   \\
\hline
Electron-Mn exchange   &$j_e$ & -0.09 meV    \\
\hline
Electron-Hole& $j_{eh}$& -0.73 meV \\
\hline
Uniaxial Anisotropy        & $D$& 10 $\mu eV$ \\
\hline
In plane  Anisotropy        & $E$& 0  \\
\hline
Quantum dot width        & $L_y$& 6nm  \\
\hline
Quantum dot width        & $L_x$& 5nm  \\
\hline
Quantum dot height       & $L_z$& 3nm  \\
\hline \hline
\end{tabular}
\end{center}
\caption{Parameters used in the simulation of the resonant
PL observed in the time resolved optical pumping
experiments.} \label{parameters }
\end{table}

In figure (\ref{fig7}) we show the result of our
simulations for a dot at thermal equilibrium ($k_BT=4$K) at
$t=0$ which is pumped with a laser pulse resonant with the
$X=-1,M_z=-\frac{5}{2}$ transition, which is the high energy one,
since the hole is parallel to Mn spin.  The laser pulse has
a duration of 300 nanoseconds, so that the spectral
broadening is negligible.  In the upper panel we plot the
PL coming from the counter-polarized transition,
$X=+1,M_z=-\frac{5}{2}$, which has lower energy and can be detected
without interference with the laser, for two different
pumping power intensity. It is apparent that after a rise
of the PL in a  time scale of 8 ns, corresponding the spin
relaxation of the exciton spin from $X=-1$ to $X=+1$, the
PL signal is depleted. The origin of the depletion is seen
in figure~(\ref{fig9}). The occupation of the $M_z=-\frac{5}{2}$
spin state in the ground reduced down to zero, in benefit
of the other Mn spin states.

Accordingly, the average magnetization becomes finite. Thus,
net angular momentum is transferred from the laser to the
Mn spin. The transfer takes place through Mn spin
relaxation enabled in the presence of the exciton. As
discussed above, the most efficient mechanism combines
hole-spin relaxation due to phonons combined with
dark-bright mixing, which involves a Mn spin flip.

Interestingly, the fact that in the steady state several Mn
spin states are occupied, including the higher energy ones,
is compatible with a picture in which the Mn spin is
precessing. Thus, a steady supply of spin-polarized
excitons in the dot would result in the precession of the
Mn spin, a scenario  similar to that of current drive
spin-torque oscillators\cite{Kiselev03}. Further work
necessary to confirm this scenario is outside the scope of
this paper.

\begin{figure}[t]
\begin{center}
\includegraphics[angle=0,width=1\linewidth]{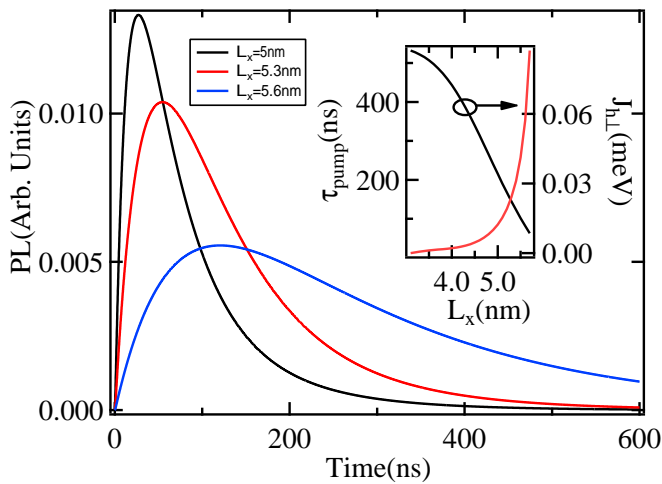}
\end{center}
\caption{ \label{fig11}(Color online) Evolution of Mn spin
orientation efficiency as a function of the valence band
mixing controlled by the anisotropy of the confinement
potentiel (L$_y$=13nm, variable L$_x$). The pumping is
detected in the PL intensity from state ($X=+1$,$M_z=-\frac{5}{2}$)
under the influence of a driving laser pumping the system
from optical ground state $M_z=-\frac{5}{2}$ to the excited state
($X=-1$, $M_z=-\frac{5}{2}$). The inset shows the evolution of
$\tau_{pump}$ with L$_x$ and J$_{h\perp}$. The exciton
generation rate is fixed at 1 ns$^{-1}$.}
\end{figure}

The efficiency of the process increases with the laser
power, as shown in figure (\ref{fig10}).  We define the
spin orientation time $\tau_{pump}$ as the time at which
the PL of the counter polarized transition is half the
maximum. We can see that, as expected, $\tau_{pump}$ is a
decreasing function of the laser power. A pumping time
$\tau_{pump}\simeq$90ns is obtained with a generation rate of
about 1ns$^{-1}$. The amplitude of the valence band
mixing, controlled by the anisotropy of the confinement
potential or the in-plane strain distribution, is the main
quantum dot parameter controlling the efficiency of the
optical pumping. As presented in figure (\ref{fig11}),
decreasing the quantum dot anisotropy, i.e., decreasing the
LH-HH mixing parameter J$_{h\perp}$, produces a rapid
increase of $\tau_{pump}$ (inset of figure (\ref{fig11}).
This is a direct consequence of the reduction of the phonon
induced hole spin flip.

\section{Summary  and conclusions}

We have studied the spin dynamics of a single Mn atom in a
CdTe quantum dot excited by a laser that drives the
transition between the 6 optical ground states,  associated
to the $2S+1$ states of the Mn spin $S=\frac{5}{2}$,  and the  24
single exciton states, corresponding to $X=\pm 1,\pm2$
states interacting with the Mn spin. The main goal is to
have a microscopic theory for the Mn spin relaxation
mechanisms that makes it possible to produce laser induced
Mn spin orientation in a time scale of less than 100 ns.
\cite{Le-Gall_PRL_2009,Goryca_PRL_2009,Le-Gall_PRB_2010}
For that matter, we need to describe how the Mn and the
quantum dot exciton affect each other.

In section (\ref{hamil0}) we describe the different terms
in the Mn spin Hamiltonian, including exchange with the
0-dimensional exciton. The symmetry of the exchange
interaction depends on the spin properties of the carriers,
which in the case of holes are strongly affected by the
interplay of confinement, strain and spin orbit coupling.
In section (\ref{hamil0})  we also use a model for
holes\cite{dot-holes,Fernandez-Rossier_PRB_2006,Kuhn09} in
quantum dots, which permits to obtain analytical
expressions for the wave functions of the holes, the
hole-Mn exchange, in terms of the dimensions of the dot and
the Kohh-Luttinger Hamiltonian.

In section (\ref{Mn-phonon}) we study the dissipative
dynamics of the Mn spin due to its coupling to phonons,
both with and without excitons in the dot.  The Mn
spin-phonon coupling arises from the time dependent
stochastic fluctuations of the crystal field and thereby of
the single ion magnetic anisotropy, induced by the phonon
field.  Whereas the Mn spin relaxation is accelerated by 2
or 3 orders of magnitude in the presence of the exciton,
the efficiency of this mechanism is too low to account for
the optical orientation of the Mn spin reported
experimentally\cite{Le-Gall_PRL_2009,Goryca_PRL_2009,Le-Gall_PRB_2010}.
The small Mn spin-phonon coupling comes from the small
magnetic anisotropy of  Mn as a substituional impurity in
CdTe.

In section (\ref{hole-spin}) we describe the interaction
between the hole spin and the phonons in non-magnetic dots.
Using the simple analytical model for the holes presented
in section \ref{hamil0} we obtain analytical formulas for
the hole spin relaxation.  We find that hole spin lifetime
can be in the range of 30 ns for a hole spin splitting as
large as that provided by the hole-Mn coupling. Thus, we
expect that bright excitons will relax into dark excitons
via hole-spin relaxation.  This provides a microscopic
mechanism to the scenario for Mn spin relaxation  proposed
by  Cywinski \cite{Cywinski10}: bright excitons relax into
dark excitons, via carrier spin relaxation, and the joint
process of  Mn-carrier spin exchange  couples  the dark
excitons to the bright excitons, resulting in PL from dark
states which implies Mn spin relaxation in a time scale of
a few nanoseconds. This scenario is confirmed by
calculations presented in section
\ref{hole-phonon-magnetic}. Finally, in section
\ref{dynamics} we present the master equation that governs
the dynamics of the 30 states of the dot, we solve it
numerically  and we model the optical Mn spin orientation
reported experimentally.

Our main conclusions are:
\begin{itemize}
\item Mn spin-phonon spin relaxation is presumably too weak to account for Mn spin dynamics in the presence of the exciton
\item The Mn spin orientation is possible in a time scale of one hundred  nanoseconds via a combination of phonon-induced hole spin relaxation and the subsequent recombination of the dark exciton enabled by spin-flip exchange of the Mn and the carrier
\item The critical property that governs the hole-Mn exchange and the hole spin relaxation is the mixing between light and heavy holes, which depends both on the shape of the dot and on strain.
\item Our microscopic model permits to account for the optically induced Mn spin
orientation.
\end{itemize}

Future work should address how the coupling of the
electronic Mn spin to the nuclear spin modifies our
results. This probably plays a role for the Mn in the dot
without excitons.  In addition, future work should study
the role played by Mn spin coherence, and the interplay
between optical and spin coherence.

\begin{acknowledgments}
We thank F. Delgado, C.  Le Gall, R. Kolodka and H.
Mariette for fruitful discussions. This work has been
financially supported by MEC-Spain (Grants MAT07-67845,
FIS2010-21883-C02-01, and CONSOLIDER CSD2007-00010),
Generalitat Valenciana (ACOMP/2010/070), Fondation
NanoScience (RTRA Genoble) and French ANR contract QuAMOS.
\end{acknowledgments}

\appendix

\section{Kohn Luttinger Hamiltonian}
\label{KL}

The Kohn-Luttinger Hamiltonian for the 4 topmost valence bands of a Zinc Blend compound are given by:
\begin{widetext}

\begin{eqnarray}
{\cal H}(k_x,k_y,k_z)=\left(\begin{array}{cccc}
P+ Q -\frac{3}{2}\kappa\nu_B B & S &R&  0 \\
S^{\dagger} & P-Q- \frac{1}{2}\kappa \mu_B B &0 & R \\
R^{\dagger} & 0 & P-Q+\frac{1}{2}\kappa \mu_B B & -S \\
0 & R^{\dagger} & -S^{\dagger} & P+ Q +\frac{3}{2}\kappa\nu_B B
 \end{array}\right)
 \label{KL}
\end{eqnarray}
\end{widetext}
where\cite{Broido-Sham}

\begin{equation}
P=\hbar^2 \gamma_1 \frac{k_z^2+k^2_{\perp}}{2m_0} \,\,\,
Q=\hbar^2 \gamma_1 \frac{-2 k_z^2+k^2_{\perp}}{2m_0}
\end{equation}



\begin{equation}
S=2 \sqrt{3} \gamma_2 \frac{\hbar^2 k_z k_{\perp}}{2m_0}
\end{equation}
and

\begin{equation}
R=-\frac{\sqrt{3}\hbar^2}{2m_0}  \left(-\overline{\gamma} k_{-}^2+\mu k_+^2\right)
\end{equation}

where $\gamma_{1,2,3}$ are dimensionless material dependent
parameters,
$\overline{\gamma}=\frac{1}{2}(\gamma_2+\gamma_3)$,
$\mu=\frac{1}{2}(\gamma_2-\gamma_3)$, $m_0$ is the free
electron mass, $k_{\perp}^2=k_x^2+k_y^2$ and $k_{\pm}=k_x
\pm i k_y$. For the dot states the relevant parameters are:

\begin{equation}
\overline{P}=\frac{\hbar^2}{2m_0} \gamma_1 \pi^2
\left(\frac{1}{L_z^2}+\frac{1}{L_x^2}+\frac{1}{L_y^2}\right)
\end{equation}

\begin{equation}
\overline{Q}=\frac{\hbar^2 \gamma_1}{2m_0}  \pi^2  \left(\frac{-2}{L_z^2}+\frac{1}{L_x^2}+\frac{1}{L_y^2}\right)
\end{equation}

\begin{equation}
\overline{R}=-\frac{\hbar^2\pi^2}{2m_0} \sqrt{3} \gamma_2 \left(\frac{1}{L_x^2}-\frac{1}{L_y^2}\right)
\end{equation}

\section{General formula for phonon-induced spin-flip rate}
\label{phonon-app}

In this appendix we derive a general formula for the
scattering rate  between two electronic state $n$ and  $n'$
induced by a phonon emission.  The Hamiltonian of the
system can be split in 3 parts, the electronic states $n$,
the phonon states, and their mutual coupling. The phonon
states are labelled according to their polarization and
momentum, $\lambda$, $\vec{q}$.  We consider the following
coupling
 \begin{equation}
 {\cal V}= \sum_{m,m',\vec{q},\lambda} {\cal V}^{m,m'}_{\vec{q},\lambda} |m\rangle\langle m'| \left(b_{\lambda q}^{\dagger}+b_{\lambda,-q}\right)
 \label{general-coupling}
 \end{equation}
where $m$ and $m'$ are electronic states.  We refer to the
free phonon states as the reservoir states. Within the
Born-Markov approximation, the scattering rate between
states $n$ and $n'$ .
 \begin{widetext}
 \begin{equation}
\Gamma_{n\rightarrow n'} = \frac{2\pi}{\hbar}\sum_r P_r \sum_{r'} \left|\langle nr |V|n'r'\rangle\right|^2
\delta\left(E_n-E_{n'}+e_r-e_{r'}\right)
\end{equation}
\end{widetext}
where  $P_r$ is the occupation of the $r$ reservoir state
with energy $e_r$. This equation can be interpreted as a
statistical average over reservoir initial states $r$ of
the Fermi Golden rule decay rate of state $N,r$.

The sums over $r$ and $r'$ are performed using the following trick. For a given $r$, the initial reservoir state,
$r'$ must have an additional phonon, since we consider  the phonon emission case. Thus,  we write:
\begin{equation}
|r'\rangle=\frac{1}{\sqrt{n_{\lambda',q'}+1}} b^{\dagger}_{\lambda',q'} |r\rangle
\end{equation}
so that
\begin{equation}
\langle r|b^{\dagger}_{q,\lambda} + b_{-q,\lambda} |r'\rangle =
\delta_{-q,q'}\delta_{\lambda,\lambda'} \sqrt{n_{\lambda',q'}^r+1}
 \end{equation}
The matrix element
  \begin{equation}
\langle nr |V|n'r'\rangle={\cal V}^{n,n'}_{\vec{q},\lambda} \sqrt{n_{\lambda',q'}^r+1}
\end{equation}
We see how from all the terms in the sum that defines the
coupling, only one survives and fixes the index $r'$. Thus,
the only the sums left are the over the initial reservoir
states and the $\lambda,q$ index that define the final
state. Now we use the definition of the Bose function:
 \begin{equation}
\sum_r P_r
\left(n_{\lambda',q'}^r+1\right)= n_B\left(\omega_{\lambda'}(q')\right)+1
\end{equation}
  and we arrive to the following expression for the rate:
 \begin{widetext}
 \begin{equation}
\Gamma_{n\rightarrow n'} = \frac{2\pi}{\hbar}
 \sum_{\lambda,q}|{\cal V}^{n,n'}_{\vec{q},\lambda}|^2
\left(n_B\left(\omega_{\lambda}(q)\right)+1\right)
\delta\left(E_n-E_{n'}-\omega_{\lambda}(q)\right)
\label{general-rate}
\end{equation}
\end{widetext}
Notice that it is possible to write the rate as a sum over
different contributions arising from different
polarizations, $\Gamma=\sum_{\lambda} \Gamma_{\lambda}$. In
the particular case that we can neglect the dependence of
the matrix element ${\cal V}_{\vec{q},\lambda}(n,n')\simeq
|{\cal V}_{(n,n')}|^2$ on $\vec{q}$ and $\lambda$, we
arrive to the following expression:
\begin{equation}
\Gamma_{n\rightarrow n'} = \frac{2\pi}{\hbar} \left(n_B\left(\Delta\right)+1\right)
 \rho_{\lambda}(\Delta)
  |{\cal V}_{n,n'}|^2
\end{equation}
where $\rho(\Delta)\equiv \sum_{\lambda,q}
\delta\left(\Delta-\omega_{\lambda}(q)\right)$ is the density of states of the phonons evaluated at the transition energy $\Delta$.





\end{document}